\documentclass[aps,twocolumn,superscriptaddress]{revtex4}

\usepackage{bm}
\usepackage{amsmath}
\usepackage{color,multirow}

\usepackage{graphicx}
\begin{document}
\title{Superconductivity in doped Dirac semimetals}

\author{Tatsuki Hashimoto}
\affiliation{Department of Applied Physics, Nagoya University, Nagoya 464-8603, Japan}
\author{Shingo Kobayashi}
\affiliation{Department of Applied Physics, Nagoya University, Nagoya 464-8603, Japan}
\affiliation{Institute for Advanced Research, Nagoya University, Nagoya 464-8601, Japan}
\author{Yukio Tanaka}
\affiliation{Department of Applied Physics, Nagoya University, Nagoya 464-8603, Japan}
\author{Masatoshi Sato}
\affiliation{Yukawa Institute for Theoretical Physics, Kyoto University, Kyoto 606-8502, Japan}
\date{\today}

\begin{abstract}
We theoretically study intrinsic superconductivity in doped Dirac semimetals. 
Dirac semimetals host bulk Dirac points, which are formed by doubly degenerate bands, so the Hamiltonian is described by a $4 \times 4$ matrix and six types of $k$-independent pair potentials are allowed by the Fermi-Dirac statistics. 
We show that the unique spin-orbit coupling leads to characteristic superconducting gap structures and $d$ vectors on the Fermi surface and the electron-electron interaction between intra and interorbitals gives a novel phase diagram of superconductivity. It is found that when the inter-orbital attraction is dominant, an unconventional superconducting state with point nodes appears. To verify the experimental signature of possible superconducting states, we calculate the temperature dependence of bulk physical properties such as electronic specific heat and spin susceptibility and surface state. In the unconventional superconducting phase, either dispersive or flat Andreev bound states appear between point nodes, which leads to double peaks or single peak in the surface density of states, respectively.  As a result, possible superconducting states can be distinguished by combining bulk and surface measurements.
\end{abstract}

\pacs{pacs}

\maketitle
\section{Introduction}\label{sec_intro}
Unconventional superconductivity is one of the main topics in condensed matter physics. In the last decade, it has been revealed that surface states, called surface Andreev bound states (SABSs), and nodal structures in unconventional superconductors can be characterized by topological numbers of a bulk wave function \cite{PhysRevB.79.214526, PhysRevB.81.220504, PhysRevB.83.224511, PhysRevB.84.020501,PhysRevB.84.060504, 0953-8984-27-24-243201, :/content/aip/proceeding/aipcp/10.1063/1.3149495, 1367-2630-12-6-065010, PhysRevB.78.195125, PhysRevB.61.10267,doi:10.1143/JPSJ.81.011013,0034-4885-75-7-076501,doi:10.7566/JPSJ.85.072001}.

The concept of topology has also been expanding widely in the normal state since the discovery of topological insulators (TIs) having a surface Dirac cone protected by time-reversal symmetry \cite{RevModPhys.82.3045,RevModPhys.83.1057,doi:10.7566/JPSJ.82.102001}. Beside TIs, topological crystalline insulators \cite{PhysRevLett.106.106802,NatCommFu}, whose surface Dirac cones are protected by point-group symmetry instead of time-reversal symmetry, and Weyl semimetals \cite{1367-2630-9-9-356,PhysRevB.83.205101,PhysRevLett.107.186806,PhysRevLett.107.127205}, which have the bulk Weyl cones, have generated great interest owing to their outstanding electronic properties and potential applications in electronic devices \cite{PhysRevB.90.161108,TCI_app,TCI_app_strain}. 

Recent theoretical studies have revealed that the doped topological materials can be promising candidates to realize unconventional superconductivity due to their unique spin-orbit interaction \cite{PhysRevLett.105.097001,PhysRevLett.115.187001} and robustness against the nonmagnetic impurities \cite{PhysRevLett.109.187003,PhysRevB.91.060502,PhysRevB.89.214506,PhysRevB.89.155140}. In particular, the superconductivity in the TIs has been studied a lot since the observation of zero-bias conductance peak suggesting the existence of SABSs in Cu$_x$Bi$_2$Se$_3$ \cite{PhysRevLett.107.217001,Hor,doi:10.7566/JPSJ.82.044704,Bay,Zocher,PhysRevB.85.180509,PhysRevB.83.134516,PhysRevLett.108.107005,Sasaki2015206,Nagai1,Yip,0953-2048-27-10-104002,doi:10.7566/JPSJ.83.064705,Nagai3,PhysRevB.91.060502,PhysRevB.89.214506,PhysRevLett.105.097001,Kriener1,PhysRevB.90.184512,PhysRevB.90.220504,PhysRevB.90.184516,PhysRevLett.110.117001,PhysRevB.92.100508,2015arXiv151207086M}. Superconductivity in topological crystalline insulators has also been observed \cite{InSnTe_SC}, and it has been predicted that exotic SABSs appear if fully gapped odd-parity superconductivity is realized \cite{PhysRevLett.109.217004,PhysRevB.92.174527}. 
Furthermore, the realization of exotic superconductivity has been anticipated in doped Weyl semimetals \cite{PhysRevLett.114.096804,PhysRevB.92.035153,PhysRevB.86.214514}. 

In this paper, we study superconductivity in doped rotation symmetric Dirac semimetals (DSs). 
DSs are materials that host bulk Dirac cones \cite{PhysRevB.76.205304,1367-2630-9-9-356}.
Several materials have been predicted to be DSs \cite{PhysRevB.83.205101,PhysRevB.85.195320,PhysRevB.88.125427,PhysRevB.91.214517,PhysRevLett.108.140405}, and Cd$_3$As$_2$ \cite{10.1038/ncomms4786,PhysRevLett.113.027603,PhysRevB.91.241114,10.1038/srep06106,10.1038/nmat3990,10.1038/nmat4023,PhysRevLett.113.246402,10.1038/nmat4143} and Na$_3$Bi \cite{Liu21022014,Xu16012015,10.1063/1.4908158} have been confirmed experimentally. 
Recently, superconductivity has been observed in Cd$_3$As$_2$ \cite{10.1038/nmat4455,10.1038/nmat4456,2015arXiv150202509H}. Moreover, point contact experiments for Cd$_3$As$_2$ have suggested the existence of SABS \cite{10.1038/nmat4455,10.1038/nmat4456}, gathering great attention as a candidate of unconventional superconductor. In addition, two of the authors have revealed that a unique orbital texture in DSs suggests unconventional pairings ($\Delta_{2}$ and $\Delta_{3}$ in this paper) \cite{PhysRevLett.115.187001}. 

However, the physical properties of the superconducting states in doped DSs have not been examined systematically, and thus it has been difficult to identify the pairing symmetry experimentally. One of our purposes of this paper is to clarify the physical property and topological structure of possible superconducting states in doped DSs. 
Due to the presence of time-reversal symmetry and inversion symmetry, the electronic states near the Dirac points are minimally described by a 4 $\times$ 4 Dirac Hamiltonian with spin and orbital degrees of freedom. 
For the superconducting state, doubly degenerate bands allow six types of $k$-independent Cooper pairs and the unique orbital texture favors an equal-spin pairing, giving rise to point nodes on the Fermi surface. From this view point, the superconductivity in doped DSs can be unconventional and its physical implication deserves further exploration. 
Superconductivity in Dirac systems has also been studied for doped TIs, Weyl semimetals, and bilayer Rashba systems. These superconducting states also show unconventional superconductivity, but we emphasize that the crystal symmetry and spin-orbit coupling of DSs are different from them. Hence the Cooper pairs respect different irreducible representations, implying that they can show unique superconducting gap structures and $d$ vectors on the Fermi surface. 

\begin{figure}[t]
\begin{center}
\includegraphics[width=8cm]{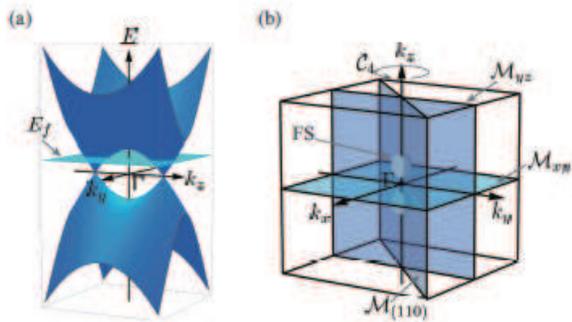}
\caption{(a) The energy dispersion of the Dirac semimetal. The bulk Dirac cones appear on the $k_z$ axis. (b) The double Fermi surface of the doped Dirac semimetal. 
Symmetry axis and plane focused in this paper are also shown.}
\label{fig_normal}
\end{center}
\end{figure}

 To clarify these points, we derive an analytical formula of possible pair potentials in the band basis and reveal the superconducting gap structure and $d$ vector on the Fermi surface. It is found that the superconducting gap structure can be classified into four types, i.e., isotropic full gap, point node at poles, horizontal line node, and vertical line node. We also show that these superconducting gap structure can be interpreted from the orbital structure of the DS and possible pair potentials. Moreover, for the odd-parity pairings, the direction of the $d$ vector is either parallel to $x$-$y$ plane or parallel to $z$-direction. These characteristics of the superconducting states are completely different from those of other topological materials. We also solve the linearized gap equation to make a superconducting phase diagram, in which an unconventional superconducting state ($\Delta_{2}$ or $\Delta_{3}$ in this paper) is realized when the inter-orbital attraction is sufficiently stronger than the intra-orbital one. To examine whether we can distinguish the possible pair potentials experimentally, we calculate the temperature dependence of the specific heat that reflects the superconducting gap structure and spin susceptibility that reflects the Van-Vleck effect and the direction of the $d$ vector. Furthermore, we calculate the surface state by using the recursive Green's function method. The unconventional superconducting states show either dispersive or flat Andreev bound states on the surface depending on the parity of mirror-reflection symmetry. Using mirror-reflection symmetry, we discuss topological numbers relevant to zero-energy states. As a result, these physical implications conclude that the possible superconducting states can be distinguished by combining bulk and surface measurements. 

This paper is organized as follows. First, we introduce a model Hamiltonian for DSs and consider possible pair potentials in Sec. \ref{sec_model}. In Sec. \ref{sec_gap}, by transforming the pair potentials from the orbital basis to the band basis, we obtain a single-band description of the pair potentials. We also show the superconducting gap structure and $d$ vector on the Fermi surface. In Sec. \ref{sec_orb}, we show that the superconducting gap structure can be interpreted from the orbital structure of DSs.
In Sec. \ref{sec_phase}, we obtain the phase diagram for the superconducting state. Numerical results for the bulk and surface states are shown in Secs. \ref{sec_bulk} and  \ref{sec_surface}, respectively. In Sec. \ref{sec_disc}, we discuss the difference between superconductivity in DSs and TIs and briefly mention superconductivity in other classes of Dirac semimetals. Finally, we summarize our results in Sec. \ref{sec_summary}.

\section{Model}\label{sec_model}
DSs have both time-reversal symmetry and inversion symmetry, which lead to doubly degenerate bands. Thus, to construct a model with four-fold degenerate Dirac points, it is necessary to take into account the orbital degrees of freedom in addition to the spin degrees of freedom. 
 In the broad sense, there are two types of DS, i.e., accidental ones and symmetry protected ones. The former type appears just on the topological phase transition point between a TI and a normal insulator \cite{1367-2630-9-9-356,PhysRevB.76.205304, PhysRevB.87.155105, PhysRevLett.109.186804}. In this case, the bulk Dirac cones are easily gapped out. On the other hand, the latter type of DSs host bulk Dirac points protected by rotational symmetry on the rotational axis and topological surface states \cite{10.1038/ncomms5898,PhysRevB.92.165120}. 
A representative example of the latter type is Cd$_3$As$_2$, where the relevant orbitals are $S_{J=\frac{1}{2}}$ and $P_{J=\frac{3}{2}}$ with a total angular momentum $J$, the and bulk Dirac points are protected by the four fold symmetry. We discuss in the following the Cd$_3$As$_2$ class DSs. In the basis set of $|s, \uparrow\rangle$, $|p_x+ip_y, \uparrow\rangle$, $|s,\downarrow\rangle$, and $|p_x-ip_y,\downarrow\rangle$, the low energy effective Hamiltonian for the DSs near $\Gamma$ point is described by
\begin{align}
H_n({\bm k})&=a(\bm{k})\sigma_z s_0+b(\bm{k})\sigma_x s_z
+c(\bm{k})\sigma_y s_0\nonumber
\\
&+d(\bm{k})\sigma_x s_x+e(\bm{k})\sigma_x s_y,
\label{H_normal}
\end{align}
where $s_i$ and $\sigma_i$ $(i=0,x,y,z)$ are the Pauli matrices in the spin and orbital space, respectively \cite{PhysRevB.88.125427,10.1038/ncomms5898}. As summarized in Ref. \cite{10.1038/ncomms5898}, the basis functions for four fold symmetric DSs are given as
\begin{align}
a(\bm{k})&=m_0-m_1 k_z^2-m_{2}(k_x^2+k_y^2),\label{eq_a}\\
b(\bm{k})&=\eta k_x,\\
c(\bm{k})&=-\eta k_y,\\
d(\bm{k})&=(\beta + \gamma) k_z (k_y^2 - k_x^2),\\
e(\bm{k})&=-2(\beta - \gamma) k_z k_x k_y,\label{eq_e}
\end{align}
where $m_0$, $m_1$, $m_2$, $\eta$, $\beta$ and $\gamma$ are material dependent parameters. The energy dispersion is shown in Fig. \ref{fig_normal} (a). By tuning the chemical potential, double Fermi surfaces appear as shown in Fig. \ref{fig_normal} (b). 

The crystals of Cd$_3$As$_2$ belong to the $D_{4h}$ point group and thus the Hamiltonian satisfies the following symmetries: 
(i) time-reversal symmetry: ${\cal T}=i\sigma_0s_y\cal K$,
\begin{align}
{\cal T}H_n({\bm k}){\cal T}^\dagger=H_n(-{\bm k});
\end{align}
(ii) inversion symmetry: ${\cal P}=\sigma_zs_0$,
\begin{align}
{\cal P}H_n({\bm k}){\cal P}^\dagger=H_n(-{\bm k});
\end{align}
(iii) four-fold rotational symmetry along $z$ axis: ${\cal C}_4=e^{i(\pi/4)(2\sigma_0+\sigma_z)s_z}$,
\begin{align}
{\cal C}_4H_n(k_x,k_y,k_z){\cal C}_4^\dagger=H_n(k_y,-k_x,k_z);
\end{align}
(iv) $x$-$y$ mirror-reflection symmetry: ${\cal M}_{xy}=i\sigma_0s_z$,
\begin{align}
{\cal M}_{xy}H_n(k_x,k_y,k_z){\cal M}_{xy}^\dagger=H_n(k_x,k_y,-k_z);
\end{align}
(v) $y$-$z$ mirror-reflection symmetry: ${\cal M}_{yz}=i\sigma_0s_x$,
\begin{align}
{\cal M}_{yz}H_n(k_x,k_y,k_z){\cal M}_{yz}^\dagger=H_n(-k_x,k_y,k_z);
\end{align}
(vi) (110) mirror-reflection symmetry: ${\cal M}_{(110)}=(\sigma_z s_x-i\sigma_0 s_0)/\sqrt{2}$,
\begin{align}
{\cal M}_{(110)}H_n(k_x,k_y,k_z){\cal M}_{(110)}^\dagger=H_n(k_y,k_x,k_z);
\end{align}
The corresponding symmetry axis and planes are shown in Fig. \ref{fig_normal} (b).

Next, we consider the superconducting state. We assume the following pair interaction \cite{PhysRevLett.105.097001}:
\begin{align}
H_{\rm int}({\bm x})=-U[n_1^2({\bm x})+n_2^2({\bm x})]-2Vn_1({\bm x})n_2({\bm x}),
\end{align}
where $U$ and $V$ are intra- and interorbital interactions, respectively, and $n_i$ ($i=1,2$) is the density operator for orbital $i$.
Then we construct the Bogoliubov de Gennes (BdG) Hamiltonian in the mean-field regime:
\begin{align}
H_{MF}=\int dk \hat c^\dagger H_{\rm BdG}({\bm k})\hat c,\\
H_{\rm BdG}({\bm k})=[H_{n}({\bm k})-\mu]\tau_z+\Delta_i\tau_x,
\label{BdG}
\end{align}
where $\tau_x$ and $\tau_z$ are the Pauli matrices in the Nambu (particle-hole) space, $\mu$ and $\Delta_i$ denote the chemical potential and pair potential, respectively. Here, the basis is taken as $\hat c^\dagger=(c^\dagger_{1\uparrow},c^\dagger_{2\uparrow},c^\dagger_{1\downarrow},c^\dagger_{2\downarrow},-c_{1\downarrow},-c_{2\downarrow},c_{1\uparrow},c_{2\uparrow})$.
Then, we discuss possible pair potentials. For the two orbital system, there are sixteen combinations of two Pauli matrices, $s_i$ and $\sigma_i$ $(i=0, x, y, z$), but six combinations out of them satisfy the Fermi-Dirac statistics, which are described by $\Delta\sigma_0s_0\equiv\Delta_{1a}$, $\Delta\sigma_zs_0\equiv\Delta_{1b}$, $\Delta\sigma_ys_y\equiv\Delta_{2}$, $\Delta\sigma_ys_x\equiv\Delta_{3}$, $\Delta\sigma_xs_0\equiv\Delta_{4a}$, and $\Delta\sigma_ys_z\equiv\Delta_{4b}$. 
These pair potentials can be classified into inter- or intraorbital in addition to the spin-singlet or triplet classes. 
 $\Delta_{1a}$ and $\Delta_{1b}$ are spin-singlet intraoribtal pairings. $\Delta_{2}$, $\Delta_{3}$, and $\Delta_{4b}$ are spin-triplet interoribtal pairings. $\Delta_{4a}$ is a spin-singlet inter-oribtal pairing. 
Moreover, these pair potentials are classified into four irreducible representations of the $D_{4h}$ point group : $A_{1g}$ ($\Delta_{1a}$ and  $\Delta_{1b}$), $B_{1u}$ ($\Delta_{2}$), $B_{2u}$ ($\Delta_{3}$) and $E_{u}$ ($\Delta_{4a}$ and $\Delta_{4b}$), which are summarized in Table \ref{table_pair}.
Symmetry properties of the pair potentials under the inversion ${\cal P}$, four-fold rotation ${\cal C}_4$ , and mirror-reflection symmetry ${\cal M}_{xy}$, ${\cal M}_{yz}$ and ${\cal M}_{(110)}$ are also summarized in Table \ref{table_pair}. As long as we consider the $k$-independent pair potentials in this orbital basis, the parity of the intra- (inter-) orbital pair potentials is even (odd) under the inversion operation. 
It is noted that the matrix forms of the possible pair potentials are common in two-orbital or layer systems such as TIs \cite{PhysRevLett.105.097001}, Weyl semimetals \cite{PhysRevB.92.035153}, and bilayer Rashba systems \cite{PhysRevLett.108.147003}. However, the superconductivity in the DSs is completely different from that in other materials since the normal state is different, as we see below. 
\begin{table}[t]
\begin{center}
\begin{tabular}{cccccccccccc}
\hline\hline
$\Delta$&&spin&orb.&Rep.&$\cal P$&${\cal{C}}_4$&${\cal{M}}_{xy}$&${\cal{M}}_{yz}$&${\cal{M}}_{zx}$&${\cal{M}}_{(110)}$\\
\hline
$\Delta_{1a}$&$\sigma_0s_0$&singlet&intra&$A_{1g}$&$+$&$+$&$+$&$+$&$+$&$+$\\
$\Delta_{1b}$&$\sigma_zs_0$&singlet&intra&$A_{1g}$&$+$&$+$&$+$&$+$&$+$&$+$\\
$\Delta_{2}$&$\sigma_ys_y$&triplet&inter&$B_{1u}$&$-$&$-$&$-$&$-$&$-$&$+$\\
$\Delta_{3}$&$\sigma_ys_x$&triplet&inter&$B_{2u}$&$-$&$-$&$-$&$+$&$+$&$-$\\
$\Delta_{4a}$&$\sigma_xs_0$&singlet&inter&$E_{u}$&$-$&$\Delta_{4b}$&$+$&$+$&$-$&$-\Delta_{4b}$\\
$\Delta_{4b}$&$\sigma_ys_z$&triplet&inter&$E_{u}$&$-$&$\Delta_{4a}$&$+$&$-$&$+$&$\Delta_{4a}$\\
\hline\hline
\end{tabular}
\caption{Possible pair potentials for the Dirac semimetals. Spin state, orbital state, irreducible representation and symmetry properties of each pairings are shown.}
\label{table_pair}
\end{center}
\end{table}
\section{Single band description of pair potentials : superconducting gap and spin structure}\label{sec_gap}
In this section, to understand the superconducting gap and spin structure on the Fermi surface, we derive the pair potentials in the band basis, where $H_n({\bm k})$ is diagonalized \cite{Yip}. Then, we extract the conduction or valence band components of  the pair potentials in order to obtain a single-band description. 
 First, we diagonalize the spin part of $H_n({\bm k})$. The Hamiltonian reduces to 
\begin{align}
H_{n\tilde s}({\bm k})&=a(\bm{k})\sigma_z+c(\bm{k})\sigma_y+{\tilde s}L(\bm{k})\sigma_x,
\end{align}
where $\tilde s =\pm1$ and $L(\bm{k})=\sqrt{d^2(\bm{k})+e^2(\bm{k})+b^2(\bm{k})}$.
The corresponding eigenvectors are given by 
\begin{align}
|\tilde s +\rangle&=\frac{1}{\sqrt{2}}
\begin{pmatrix}
\cos \frac{P_k}{2}
\\
\cos \frac{P_k}{2}
\left(\frac{d(\bm{k})+ie(\bm{k})}{L(\bm{k})+b(\bm{k})}\right)
\end{pmatrix},
\\
|\tilde s -\rangle&=\frac{1}{\sqrt{2}}
\begin{pmatrix}
\sin \frac{P_k}{2}
\\
-\sin \frac{P_k}{2}
\left(\frac{d(\bm{k})+ie(\bm{k})}{L(\bm{k})-b(\bm{k})}\right)
\end{pmatrix},
\end{align}
where $\cos \frac{P_k}{2}=\sqrt{1+\frac{b({\bm k})}{L({\bm k})}}$ and $\sin \frac{P_k}{2}=\sqrt{1-\frac{b({\bm k})}{L({\bm k})}}$. By using the eigenvectors, the following relations are obtained as 
\begin{align}
\tilde{s}_x&=-\frac{d(\bm{k})b(\bm{k})}{l(\bm{k})L(\bm{k})} s_x -\frac{e(\bm{k})b(\bm{k})}{l(\bm{k})L(\bm{k})} s_y 
+\frac{l(\bm{k})}{L(\bm{k})} s_z
\label{sp_sd_1},\\
\tilde{s}_y&= \frac{e(\bm{k})}{l(\bm{k})} s_x - \frac{d(\bm{k})}{l(\bm{k})} s_y
\label{sp_sd_2},\\
\tilde{s}_z&=\frac{d(\bm{k})}{L(\bm{k})}s_x + \frac{e(\bm{k})}{L(\bm{k})}s_y+\frac{b(\bm{k})}{L(\bm{k})}s_z,
\label{sp_sd_3}
\end{align}
where $l(\bm{k})=\sqrt{d^2(\bm{k})+e^2(\bm{k})}$, $\tilde{s}_i$ ($i=0,x,y,z$) are the Pauli matrices for the spin helicity basis.
Next, we diagonalize the orbital part. The eigenvalues are given by 
\begin{align}
E_{n\tilde \sigma}({\bm k})=\tilde \sigma\sqrt{a^2(\bm{k})+b^2(\bm{k})+c^2(\bm{k})+d^2(\bm{k})+e^2(\bm{k})},
\end{align}
where $\tilde \sigma=\pm1$.
The corresponding eigenvectors are 
\begin{align}
|\tilde s \pm, \tilde\sigma+\rangle&=\frac{1}{\sqrt{2}}
\begin{pmatrix}
\cos \frac{Q_k}{2}
\\
\cos \frac{Q_k}{2}
\left(\frac{\tilde s L(\bm{k})+ic(\bm{k})}{R(\bm{k})+a(\bm{k})}\right)
\end{pmatrix},
\\
|\tilde s \pm, \tilde\sigma-\rangle&=\frac{1}{\sqrt{2}}
\begin{pmatrix}
\sin \frac{Q_k}{2}
\\
-\sin \frac{Q_k}{2}
\left(\frac{\tilde s L(\bm{k})+ic(\bm{k})}{R(\bm{k})-a(\bm{k})}\right)
\end{pmatrix},
\end{align}
where $R(\bm{k})=|E_{n\tilde \sigma}(\bm k)|$, $\cos \frac{Q_k}{2}=\sqrt{1+\frac{a({\bm k})}{R({\bm k})}}$, $\sin \frac{Q_k}{2}=\sqrt{1-\frac{a({\bm k})}{R({\bm k})}}$. Finally, the eigenvectors for the normal Hamiltonian are obtained as 
\begin{align}
|u_{\tilde s,\tilde \sigma}\rangle=|\tilde s \pm\rangle\otimes|\tilde s \pm, \tilde\sigma \pm\rangle.
\end{align}
By using the eigenvectors $|u_{\tilde s,\tilde \sigma}\rangle$, we obtain the pair potentials in the band and spin helicity basis. Then, with Eqs. (\ref{sp_sd_1}) - (\ref{sp_sd_3}), we transform the pair potentials from the band and spin helicity basis to the band and real spin basis: $\hat c^{\prime\dagger}=(c^\dagger_{\alpha\uparrow},c^\dagger_{\beta\uparrow},c^\dagger_{\alpha\downarrow},c^\dagger_{\beta\downarrow},-c_{\alpha\downarrow},-c_{\beta\downarrow},c_{\alpha\uparrow},c_{\beta\uparrow})$ where $\alpha$ and $\beta$ are band indices. The obtained results are
\begin{align}
\Delta_{1a}^\prime({\bm k})
&=\Delta\tilde\sigma_0 s_0,\label{eq_d1a}\\
\Delta_{1b}^\prime({\bm k})
&=
\frac{\Delta}{R(\bm{k})}
\left(
r(\bm{k})\tilde\sigma_x+a(\bm{k})\tilde\sigma_z
\right)s_0,
\\
\Delta_{2}^\prime({\bm k})
&=
-\Delta
\Big[
\frac{a(\bm{k})}{r(\bm{k}) R(\bm{k})}\tilde\sigma_x
(b(\bm{k}) s_x+c(\bm{k}) s_y-d(\bm{k}) s_z)\nonumber\\
&+
\frac{e(\bm{k})}{r(\bm{k})}\tilde\sigma_y s_0
-
\frac{1}{R(\bm{k})}
\tilde\sigma_z
(b(\bm{k}) s_x+c(\bm{k}) s_y-d(\bm{k}) s_z)\Big],
\\
\Delta_{3}^\prime({\bm k})
&=
-\Delta\Big[
\frac{a(\bm{k})}{r(\bm{k}) R(\bm{k})}\tilde\sigma_x
(c(\bm{k}) s_x-b(\bm{k}) s_y+e(\bm{k}) s_z)\nonumber\\
&
+
\frac{d(\bm{k})}{r(\bm{k})}\tilde\sigma_y
s_0
-
\frac{1}{R(\bm{k})}
\tilde\sigma_z
(c(\bm{k}) s_x-b(\bm{k}) s_y+e(\bm{k}) s_z)\Big],
\\
\Delta_{4a}^\prime({\bm k})
&=
-\Delta\Big[
\frac{a(\bm{k})}{r(\bm{k}) R(\bm{k})}\tilde\sigma_x
\left(
d(\bm{k})s_x+e(\bm{k})s_y+b(\bm{k})s_z
\right)\nonumber\\
&
-\frac{c(\bm{k})}{r(\bm{k})}\tilde\sigma_y
s_0
-\frac{1}{R(\bm{k})}
\tilde\sigma_z
\left(
d(\bm{k})s_x+e(\bm{k})s_y+b(\bm{k})s_z
\right)\Big],
\\
\Delta_{4b}^\prime({\bm k})
&=
-\Delta\Big[\frac{a(\bm{k})}{r(\bm{k}) R(\bm{k})}\tilde\sigma_x
\left(
-e(\bm{k})s_x+d(\bm{k})s_y+c(\bm{k})s_z
\right)\nonumber\\
&
+
\frac{b(\bm{k})}{r(\bm{k})}\tilde\sigma_y
s_0
-\frac{1}{R(\bm{k})}
\tilde\sigma_z
\left(
-e(\bm{k})s_x+d(\bm{k})s_y+c(\bm{k})s_z
\right)\Big],\label{eq_d4b}
\end{align}
\begin{figure*}[t]
\begin{center}
\includegraphics[width=16cm]{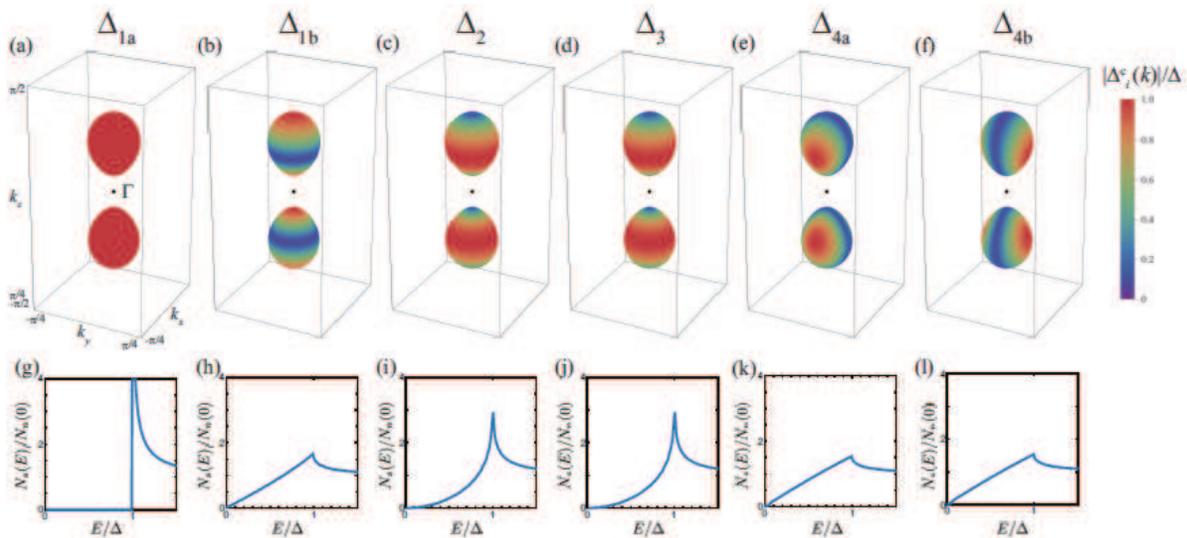}
\caption{(a) - (f) Superconducting gap structure on the Fermi surface and (g) - (l) the bulk density of states for the possible pair potentials. The color on the Fermi surface indicates the magnitude of the energy gap $|\Delta_i^c({\bm k})|/\Delta$ ($i=1a, 1b, 2, 3, 4a, 4b$).}
\label{fig_SC_gap}
\end{center}
\end{figure*}
\begin{figure*}[t]
\begin{center}
\includegraphics[width=18cm]{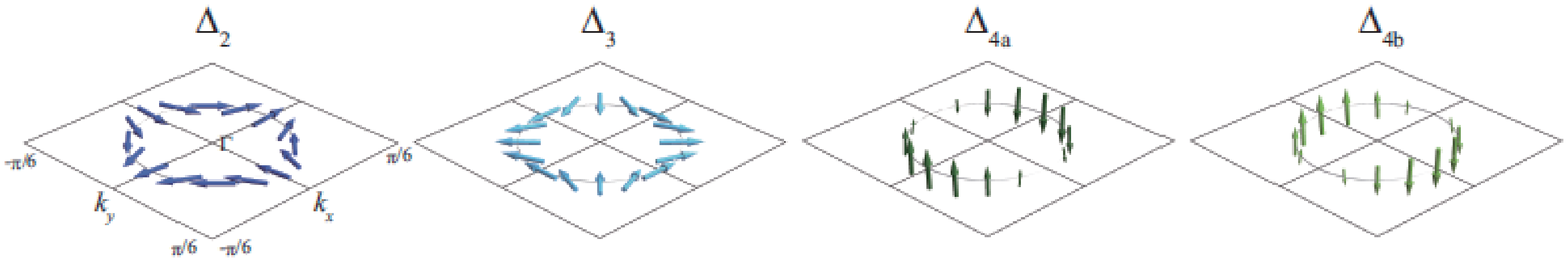}
\caption{$d$ vector on the Fermi surface for the spin-triplet pairings in the band space. In the case of $\Delta_{2}$ and $\Delta_{3}$, $d$ vectors are almost parallel to the $x$-$y$ plane. On the other hand, in the case of $\Delta_{4a}$ and $\Delta_{4b}$, the $d$ vectors are almost parallel to the $z$ direction. Here, we show the $d$ vector on the Fermi surface at $k_z=\pi/4$, however, the $k_z$ dependence of the direction of the $d$ vector is quite small since this is induced by $k^3$ terms.}
\label{fig_dvec}
\end{center}
\end{figure*}
where $\tilde\sigma_i$ $(i=0, x, y, z)$ are the Pauli matrices in the band basis and $r(\bm{k})=\sqrt{b^2(\bm{k})+c^2(\bm{k})+d^2(\bm{k})+e^2(\bm{k})}$.
In the case of the DS, the Fermi surface consists of either an electron or a hole band. If the chemical potential is large enough compared with the magnitude of the pair potential, $|\mu|>>\Delta$, we can consider that the superconductivity occurs in either conduction or valence band (quasi-classical approximation). Thus the inter-band and valence (or conduction) band component can be ignored. Namely, it is sufficient to consider only the (1,1) [or (2,2)] component of $\tilde{\sigma}_i$. The conduction band components of pair potentials are as follows:
\begin{align}
\Delta_{1a}^c({\bm k})
&=\Delta s_0,
\label{del_c_d1a}
\\
\Delta_{1b}^c({\bm k})
&=
\frac{\Delta}{R}[m_0-m_1 k_z^2-m_{2}(k_x^2+k_y^2)]s_0,
\label{del_c_d1b}
\\
\Delta_{2}^c({\bm k})
&=\Delta\Big[
\frac{\eta}{R}
(k_x s_x-k_y s_y)
+
\frac{\beta+\gamma}{R}
k_z(k_y^2-k_x^2)s_z\Big],
\label{del_c_d2a}
\\
\Delta_{3}^c({\bm k})
&=-\Delta\Big[
\frac{\eta}{R}
(k_y s_x+k_x s_y)
+2
\frac{\beta-\gamma}{R}
k_xk_yk_zs_z\Big],
\label{del_c_d2b}
\\
\Delta_{4a}^c({\bm k})
&=\Delta\Big[
\frac{\eta}{R}
k_xs_z\nonumber\\
&+
\frac{\beta+\gamma}{R}
k_z(k_y^2-k_x^2)s_x
-2
\frac{\beta-\gamma}{R}
k_xk_yk_zs_y\Big],
\label{del_c_d3a}
\\
\Delta_{4b}^c({\bm k})
&=-\Delta\Big[
\frac{\eta}{R}
k_ys_z\nonumber\\
&-2
\frac{\beta-\gamma}{R}
k_xk_yk_zs_x
-
\frac{\beta+\gamma}{R}
k_z(k_y^2-k_x^2)s_y\Big].
\label{del_c_d3b}
\end{align}
It is found that, in the single-band description, $\Delta_{1a}$ and $\Delta_{1b}$ are regarded as spin-singlet even-parity pairings, and $\Delta_{2}$, $\Delta_{3}$, $\Delta_{4a}$ and $\Delta_{4b}$ are spin-triplet odd-parity pairings. In the single-band description, the spin-singlet or triplet completely correspond to even or odd under the inversion operation for all pairings. Although $\Delta_{4a}$ is a spin-singlet inter-orbital pairing in the orbital basis, it is considered as a spin-triplet pairing in the band basis. In other words, the spin-triplet component of $\Delta_{4a}$ is induced by the spin-orbit interaction. In addition, if the parameters related to the spin-orbit interaction are absent $\eta=\beta=\gamma=0$, the odd-parity pairings $\Delta_2=\Delta_3=0$, which means that the spin-orbit interaction is essential to realize unconventional superconductivity.
As shown below, these single-band representations are useful to capture bulk superconducting properties such as the heat capacity and the spin susceptibility. 

\subsection{Superconducting gap structure}
In Fig. \ref{fig_SC_gap}, we show the magnitude of the superconducting gap $|\Delta_{i}^c({\bm k})|/\Delta$ plotted on the Fermi surface (a)-(f) and the bulk density of state (DOS) $N_s(E)/N_n(0)$ (g)-(h). 
In the case of $\Delta_{1a}$, the gap structure has an isotropic full gap, where the DOS diverges at $E/\Delta=1$ and there is no state in $E/\Delta<1$. In the case of $\Delta_{1b}$, line nodes exist in the horizontal direction. Therefore, the DOS is proportional to $E$ and divergence at $E/\Delta=1$ is suppressed. It should be noted that the line nodes are accidental nodes, and thus, by tuning some parameters, we can remove the nodes without any topological phase transition. For $\Delta_{2}$ and $\Delta_{3}$, the superconducting gap has point nodes on $k_z$-axis. The DOS near $E/\Delta=0$ is proportional to $E^2$. 
As is seen from Fig. \ref{fig_SC_gap}, the superconducting gaps of $\Delta_{2}$ and $\Delta_{3}$ are quite similar. This is because $\Delta_{2}$ and $\Delta_{3}$ are different only by $k^3$ terms, i.e., $d(\bm{k})$ and $e(\bm{k})$ in Eq. \ref{H_normal}.
For $\Delta_{4a}$ ($\Delta_{4b}$), there are point nodes on the $k_z$ axis. In the absence of the $k^3$ terms, the point nodes become the line node at $k_x=0$ ($k_y=0$) for $\Delta_{4a}$ ($\Delta_{4b}$). 
Although the $k^3$ terms change the superconducting gap structure, the superconducting gap structure of $\Delta_{4a}$ can be effectively considered as a line node, as is obvious from Figs. \ref{fig_SC_gap} (e) and \ref{fig_SC_gap} (f), since the gap opening effect of the $k^3$ terms is quite small compared with the $k$-linear terms around $\Gamma$ point. Then, the DOS is proportional to $E^2$ at very near $E/\Delta=0$ but the line shape is almost $E$ linear in the wide region of $E/\Delta<1$.
The results for the superconducting gap with and without the $k^3$ terms considered are summarized in Table \ref{table_SC_gap}.

It has been revealed that if the pair potential satisfies the four fold rotational symmetry $C_4\Delta_iC_4^t=e^{-i\pi \varphi/2}\Delta_i$ with non zero $\varphi$, the superconducting state inherits the $C_4$ invariant Dirac points of the normal state \cite{PhysRevLett.115.187001}. In the case of $\Delta_2$, $\Delta_3$, $\Delta_{4a}$ and $\Delta_{4b}$, $\varphi$ is non zero. Therefore, we can say that the point nodes on $k_z$ axis in $\Delta_2$ and $\Delta_3$ originate from the normal state, which are topologically protected.

\begin{table*}[t]
\begin{center}
\begin{tabular}{c|cc|cc}
\hline\hline
&SC gap&SC gap&$d$-vector&$d$-vector\\
&without $k^3$ terms&with $k^3$ terms&without $k^3$ terms&with $k^3$ terms\\
\hline
$\Delta_{1a}$&FG&FG&none&none\\
$\Delta_{1b}$&horizontal LNs&horizontal LNs&none&none\\
$\Delta_{2}$&PNs at poles&PNs at poles&$d\parallel$ $x$-$y$ plane&$x$-$y$ component is dominant\\
$\Delta_{3}$&PNs at poles&PNs at poles&$d\parallel$ $x$-$y$ plane&$x$-$y$ component is dominant\\
$\Delta_{4a}$&vertical LN&PNs at poles&$d\parallel$ $z$-axis&$z$-component is dominant\\
$\Delta_{4b}$&vertical LN&PNs at poles&$d\parallel$ $z$-axis&$z$-component is dominant\\
\hline\hline
\end{tabular}
\caption{Superconducting (SC) gap structure and $d$ vector for the possible pair potential. FG, PN and LN stand for full gap, point node and line node, respectively. }
\label{table_SC_gap}
\end{center}
\end{table*}
\subsection{$d$ vector}
For spin-triplet superconductors, the pair potentials can be described with $d$ vectors, which behave like three-dimensional vectors in spin space \cite{PhysRev.131.1553}. 
In our basis, the $d$ vector ${\bm d}({\bm k})$ is defined as 
\begin{align}
\Delta_i^c({\bm k})=\Delta{\bm d}({\bm k})\cdot{\bm s},
\end{align}
where ${\bm s}=(s_x, s_y, s_z)$ 
For the possible pair potentials, ${\bm d}({\bm k})$ is easily obtained from Eqs. (\ref{del_c_d1a}) - (\ref{del_c_d3b}).
The direction of the $d$ vector on the Fermi surface is important to interpret the magnetic response of the superconductivity, the details of which are mentioned in Sec. \ref{sec_bulk}.
In Fig. \ref{fig_dvec}, we show the $d$ vector of the spin-triplet pair potentials in the band basis, i.e., $\Delta_{2}$, $\Delta_{3}$, $\Delta_{4a}$ and $\Delta_{4b}$.
Note that we show the $d$ vector for $k_z=\pi/4$, but the $k_z$ dependence of the direction of the $d$-vector is negligible since it originates from the $k^3$ terms.
In the case of $\Delta_2$ and $\Delta_3$, the direction of the $d$-vector is almost parallel to the $x$-$y$ plane. Although there is an $s_z$ component, it is much smaller than the other components since it is induced by the $k^3$ terms. 
On the other hand, in the case of $\Delta_{4a}$ and $\Delta_{4b}$, the direction of the $d$ vector is almost parallel to the $z$ axis, and $x$-$y$ plane component is negligible for the same reason in the case of $\Delta_2$ and $\Delta_3$. 

\section{Interpretation of superconducting gap structure with orbital texture}\label{sec_orb}
In this section, we interpret the superconducting gap structures from the orbital texture of DSs. In the previous letter, two of the present authors have argued how the orbital texture is consistent with $\Delta_2$ and $\Delta_3$ \cite{PhysRevLett.115.187001}. Here we generalize the argument and explain the nodal structures of all possible pairing symmetries in terms of the orbit-momentum locking. For simplicity, we ignore the $k^3$ terms in this section.

First, we discuss the orbit-momentum locking of DSs. Consider the Hamiltonian in Eq. (\ref{H_normal}). The spin is already diagonalized in Eq. (\ref{H_normal}), so we can divide the Hamiltonian into spin-up and spin-down sectors as
\begin{align}
H_{n\uparrow}({\bm k})&=a(\bm{k})\sigma_z+\eta(k_x\sigma_x-k_y\sigma_y),\label{H_up}\\
H_{n\downarrow}({\bm k})&=a(\bm{k})\sigma_z-\eta(k_x\sigma_x+k_y\sigma_y).\label{H_down}
\end{align}
These Hamiltonians have two characteristic features. First, the first terms in Eqs. (\ref{H_up}) and (\ref{H_down}) dominate on the $k_z$ axis. Because of the uniaxial rotational symmetry around the $z$-direction in DSs, the orbital mixing second terms are not allowed on the $k_z$ axis. Second, the orbital mixing second terms become dominant away from the $k_z$ axis at each Dirac point. At Dirac points, both the first and the second terms vanish, but since the second terms are linear in $k_i$ while the first ones are quadratic, the second terms are dominant except on the $k_z$ axis. It should be noted that these two features are required by the symmetry of DSs.  
   
\begin{figure*}[t]
\begin{center}
\includegraphics[width=15cm]{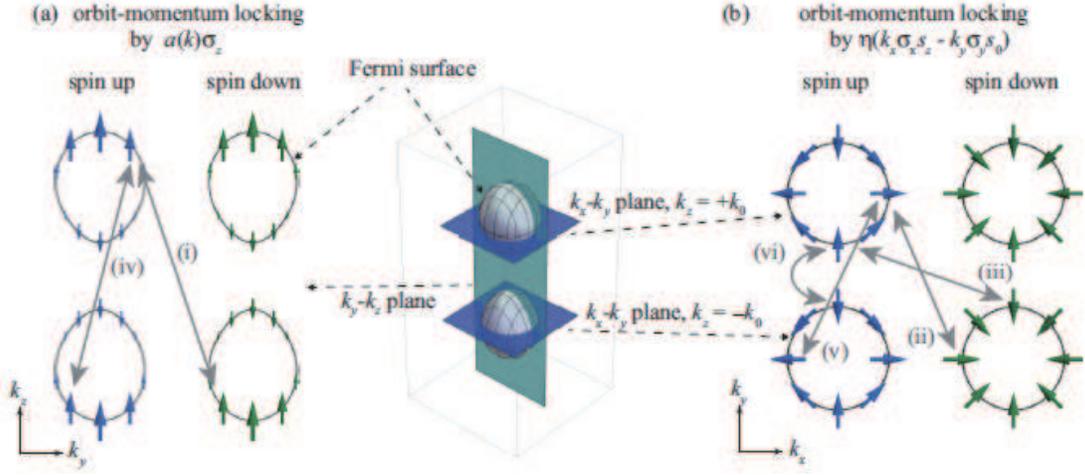}
\caption{Orbital texture of the Dirac semimetal for spin-up and spin-down space. Single-headed arrows on the Fermi surface indicate orbital ${\bm \sigma}=(\sigma_x,\sigma_y,\sigma_z)$. Orbit-momentum locking by the first term in Eq. (\ref{H_normal}): $a(k)\sigma_z$ plotted on the $k_y$-$k_z$ plane (a). Orbit-momentum locking by the second term in Eq. (\ref{H_normal}): $\eta(k_x\sigma_xs_z-k_y\sigma_ys_0)$ plotted on the $k_x$-$k_y$ plane at $k_z=\pm k_0$ (b). 
For a Cooper pair between electrons in the different spin sectors, the orbit-momentum locking provides the parallel (anti-parallel) orbital configuration in the $z$- and $x$- ($y$-) directions, which are indicated by double-headed arrows (i)  and (ii) [(iii)]. For an equal-spin Cooper pair, the orbit-momentum locking provides the parallel (anti-parallel) orbital configuration in the $z$- ($x$- and $y$-) direction, which is indicated by a double-headed arrow (iv) [(v) and (vi)].}
\label{fig_orb}
\end{center}
\end{figure*}

The above features give rise to a unique orbital texture on the Fermi surfaces surrounding the Dirac points. Near the poles of the Fermi surface, the first terms in Eqs. (\ref{H_up}) and (\ref{H_down}) are dominant, so we have the $z$-directed parallel orbital configuration shown in Fig. \ref{fig_orb} (a). On the other hand, near the equators of the Fermi surfaces, the second terms are dominant, so we have the orbit-momentum locking structure in Fig. \ref{fig_orb} (b).  

Now consider the pairing states in DSs, and compare them with the orbital textures. According to the BCS theory, Cooper pairs form between electrons with opposite momenta, $-\bm k$ and $\bm k$, and, for $\Delta_{1a}$, $\Delta_{1b}$, $\Delta_{4a}$, $\Delta_{4b}$, ($\Delta_2$, $\Delta_3$), they form between electrons in  different (same) spin sectors. First, consider a Cooper pair between electrons in the different spin sectors. As illustrated in (i) of Fig. \ref{fig_orb} (a), near the poles of the Fermi surface, the Cooper pair has a parallel orbital configuration in the $z$-direction. On the other hand, on the equator of the Fermi surface, the Cooper pair has a parallel orbital configuration in the $x$ direction [(ii) in Fig. \ref{fig_orb} (b)], or an antiparallel-orbital configuration in the $y$ direction [(iii) in Fig. \ref{fig_orb} (b)]. It is found that these orbital configurations are consistent with $\Delta_{1a}$. Note that $\sigma_0$ is diagonal on the quantization basis of $\sigma_x$ or $\sigma_z,$ but it is off diagonal on the quantization basis of $\sigma_y$ \cite{footnote1}. This means that $\Delta_{1a}$, which is proportional to $\sigma_0$, has parallel orbital configurations in $x$ and $z$ directions but an anti-parallel one in the $y$ direction, which is exactly the same as the aforementioned orbital structure of Cooper pairs in DSs. On the other hand, the other gap functions, $\Delta_{1b}$, $\Delta_{4a}$, and $\Delta_{4b}$ are not fully consistent with the orbital texture. As summarized in Table \ref{table_orb}, $\sigma_x$ ($\sigma_z$) indicates an antiparallel-orbital pair in the $z$ ($x$) direction and a parallel-orbital pair in other directions, and $\sigma_y$ indicates an antiparallel orbital pair in any direction. Therefore, for instance, $\Delta_{1b}$ is inconsistent with the orbital texture on the equator of the Fermi surfaces, so it has horizontal line nodes. In a similar manner, one can see that vertical line nodes of $\Delta_{4a}$ and $\Delta_{4b}$ come from the inconsistency between the pairings and the orbital texture of DSs. 

 For a Cooper pair between electrons in the same spin sector, the orbital texture in Fig. \ref{fig_orb} (a) gives a parallel orbital configuration in the $z$ direction near the poles of the Fermi surface [(iv)], however, that in Fig. \ref{fig_orb} (b) provides an anti-parallel orbital configuration in the $x$ and $y$ directions on the equator of the Fermi surfaces [(v),  (vi)]. Since $\sigma_y$ represents an antiparallel orbital pair in any direction, these orbital configurations are consistent with $\Delta_2$ and $\Delta_3$ on the equator of the Fermi surface, but not consistent with them near the poles. Consequently, there arise point nodes at the poles for these gap functions. 

\begin{table}[]
\begin{center}
\begin{tabular}{ccccc}
\hline\hline
&$\Delta_{1a}(\sigma_0)$&$\Delta_{1b}(\sigma_z)$&$\Delta_{4a}(\sigma_x)$&$\Delta_{2,3,4b}(\sigma_y)$\\
\hline
$x$&P&AP&P&AP\\
$y$&AP&P&P&AP\\
$z$&P&P&AP&AP\\
\hline\hline
\end{tabular}
\caption{Orbital configuration of the possible pair potentials for each direction. P (AP) stands for parallel (anti-parallel) orbital configuration.}
\label{table_orb}
\end{center}
\end{table}

\section{Phase diagram}\label{sec_phase}
In this section, we obtain the $U$-$V$ phase diagram by solving the linearized gap equation in a manner similar to the superconducting TI \cite{PhysRevLett.105.097001} and the bilayer Rashba system \cite{PhysRevLett.108.147003}. For simplicity, we ignore the $k^3$ terms in this section. The linearized gap equations for possible pair potentials are given by 
\begin{align}
   \left|
    \begin{array}{cc}
      Uq_{1a}-1 & Uq_{1ab} \\
      Uq_{1ba} & Uq_{1b}-1\\
    \end{array}
  \right|
&=0\hspace{10pt}{\rm for}\hspace{5pt}\Delta_{1},
\\
Vq_i&=1\hspace{10pt}{\rm for}\hspace{5pt}\Delta_{2},\Delta_{3},\Delta_{4a}.
\end{align}
Here, $q_i$ $(i=1a,1b,1ab,1ba,2,3,4a)$ is the irreducible susceptibility:
\begin{align}
q_i&=-\frac{k_BT}{N}\sum_{\omega_n,\bm k}{\rm Tr}\left[\frac{{\Delta}_i}{\Delta}G_0(\bm k)\frac{{\Delta}_i}{\Delta}G_0(\bm k)\right]\\
&=-\frac{1}{N}\sum_{\bm k}\left[F_i(\bm k)\frac{\tanh\frac{E_{n+}(\bm k)}{2k_BT}}{2 E_{n+}(\bm k)}\right],
\end{align}
where $N$ is the number of the unit cell, $k_B$ is the Boltzman constant, $G_0(\bm k)$ is the single-particle Green's function for the normal state, $G_0(\bm k)=P_c(\bm k)/(i\omega_n-E_{n+}(\bm k))$ with the projection operator onto the conduction band $P_c(\bm k)=\sum_{\tilde s}|u_{\tilde s,+}\rangle \langle u_{\tilde s,+}|=[E_{n+}(\bm k)\sigma_0s_0+H_n(\bm k)]/{2E_{n+}(\bm k)}$, and $F_i(\bm k)$ is the form factor originating from the orbital degrees of freedom. $F_i(\bm k)$ for each pair potential is given by $F_{1a}(\bm k)=1$, $F_{1ab}(\bm k)=F_{1ba}(\bm k)=a(\bm k)/R(\bm k)$, $F_{1b}(\bm k)=a^2(\bm k)/R^2(\bm k)$, $F_{2}(\bm k)=F_{3}(\bm k)=\eta^2(k^2_{x}+k^2_{y})/R^2(\bm k)$, and $F_{4a}(\bm k)=\eta^2k^2_{x}/R^2(\bm k)$. 

\begin{figure}[t]
\begin{center}
\includegraphics[width=5cm]{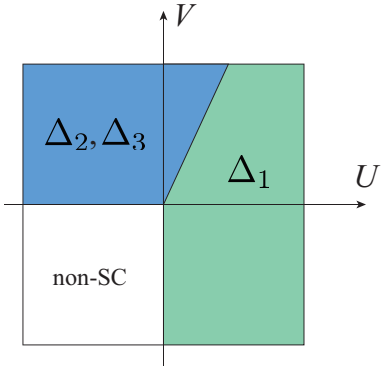}
\caption{$U$-$V$ phase diagram for the superconducting DS, where $U$ ($V$) is the inter- (intra-) orbital attraction. The blue (green) region indicates the region where the $\Delta_{2}$ and $\Delta_{3}$ ($\Delta_1$) has the highest stability.}
\label{fig_phase}
\end{center}
\end{figure}

The $U$-$V$ phase diagram is shown in Fig. \ref{fig_phase}. 
As is obvious from the form factors, the irreducible representations satisfy the relation: $q_{2}=q_{3}>q_{4a}$. Therefore, the $U$-$V$ phase diagram consists of $\Delta_1$, $\Delta_2$ and $\Delta_3$, and $\Delta_{4a}$ ($\Delta_{4b}$) cannot appear in the phase diagram. 
The phase boundary is given by 
\begin{align}
\frac{V}{U}=\int_{\rm FS}d{\bm k}\frac{R^2({\bm k})+a^2({\bm k})}{R^2({\bm k})-a^2({\bm k})}\approx 2.1.
\end{align}
If the interorbital attraction $V$ is sufficiently stronger than the intraorbital attraction $U$, the unconventional superconducting phase ($\Delta_{2}$ or $\Delta_{3}$) is realized. We can expect that the Coulomb repulsion leads to stronger $V$  as is discussed in the superconducting TI \cite{PhysRevB.90.184512}.
We can also interpret this phase diagram with the orbital structure.
Since all inter-orbital pairings in Table \ref{table_pair} are odd under parity, 
one can naturally obtain an odd-parity pairing state if $V$ dominates. 
Then, among the odd-parity pairing states in Table \ref{table_pair},  only $\Delta_2$ and $\Delta_3$ are consistent with 
the orbital texture on the equator of the Fermi surface.
Consequently, we obtain $\Delta_2$ and $\Delta_3$ in the phase diagram.
\section{Bulk physical property}\label{sec_bulk}
In this section, we obtain bulk physical properties for the possible pair potentials.
\subsection{Specific heat}
Here, we calculate the temperature dependence of the electronic specific heat for the superconducting states. The temperature dependence of the specific heat reflects the superconducting gap structure \cite{RevModPhys.63.239}, which has been considered as useful information to determine the symmetry of pair potential experimentally \cite{doi:10.7566/JPSJ.82.044704,Kriener1,PhysRevLett.71.1740, doi:10.1143/JPSJ.71.404, doi:10.1143/JPSJ.69.572}. The specific heat in the superconducting state $C_{s}$ is given by 
\begin{align}
C_{s}=\frac{2\beta}{N}\sum_{{\bm k}}&\left(E_{ij}^2({{\bm k}})+\beta E_{ij}({{\bm k}})\frac{\partial \Delta}{\partial \beta}\frac{\partial E_{ij}({{\bm k}})}{\partial \Delta}\right)\nonumber\\
&\times \left(-\frac{\partial f(E_{ij}({\bm k}))}{\partial E_{ij}({\bm k})}\right),\label{spe_eq}
\end{align}
where $\beta=1/k_BT$ and $E_{ij}$ is the eigenvalue for the BdG Hamiltonian Eq. (\ref{BdG}) given as
\begin{align}
E_{1j}(\bm{k})&=\sqrt{\xi^2(\bm{k})+2\sqrt{f^2(\bm{k})g^2(\bm{k})+\Delta^2 \zeta_j^2(\bm{k})}},\\
E_{2j}(\bm{k})&=\sqrt{\xi^2(\bm{k})-2\sqrt{f^2(\bm{k})g^2(\bm{k})+\Delta^2 \zeta_j^2(\bm{k})}},\\
E_{3j}(\bm{k})&=-\sqrt{\xi^2(\bm{k})-2\sqrt{f^2(\bm{k})g^2(\bm{k})+\Delta^2 \zeta_j^2(\bm{k})}},\\
E_{4j}(\bm{k})&=-\sqrt{\xi^2(\bm{k})+2\sqrt{f^2(\bm{k})g^2(\bm{k})+\Delta^2 \zeta_j^2(\bm{k})}}.
\end{align}
Here, 
\begin{align}
\xi^2(\bm{k})=f^2(\bm{k})+g^2(\bm{k})+\Delta^2,
\end{align}
and $\zeta_j(\bm{k})$ $(j=1a,1b,2,3,4a,4b)$ for each pair potential is
\begin{align}
\zeta_{1a}^2(\bm{k})&=0,\\
\zeta_{1b}^2(\bm{k})&=b^2(\bm{k})+c^2(\bm{k})+d^2(\bm{k})+e^2(\bm{k}),\\
\zeta_{2}^2(\bm{k})&=a^2(\bm{k})+e^2(\bm{k}),\\
\zeta_{3}^2(\bm{k})&=a^2(\bm{k})+d^2(\bm{k}),\\
\zeta_{4a}^2(\bm{k})&=a^2(\bm{k})+c^2(\bm{k}),\\
\zeta_{4b}^2(\bm{k})&=a^2(\bm{k})+b^2(\bm{k}).
\end{align}
We assume that the superconducting gap has the following phenomenological temperature dependence obtained from the BCS theory:
\begin{align}
\Delta(T)=1.76k_B T_c \tanh(1.74\sqrt{T_c/T-1})
\end{align}
where $T_c$ is the superconducting critical temperature.

\begin{figure}[t]
\begin{center}
\includegraphics[width=7cm]{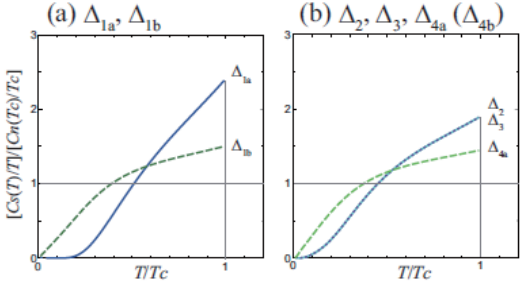}
\caption{Temperature dependence of the specific heat for (a) $\Delta_{1a}$ (blue solid line) and $\Delta_{1b}$ (green dashed line), and (b) $\Delta_{2}$ (cyan solid line), $\Delta_{3}$ (purple dotted line) and $\Delta_{4a}$ (light green dashed line). }
\label{fig_spe}
\end{center}
\end{figure}

In Fig. \ref{fig_spe}, we show the temperature dependence of the specific heat $C_s/T$ as a function of the temperature $T$. In the case of $\Delta_{1a}$,  $C_s/T$ has an exponential behavior near $T/T_c=0$, and the magnitude of the specific heat jump at $T_c$ is the largest among all possible pair potentials. 
On the other hand, in the case of $\Delta_{1b}$, the superconducting gap has line nodes, which leads to a linear behavior of $C_s/T$ around $T/T_c=0$. To satisfy the entropy balance, $\int_{0}^{T_c}dT[C_s(T)-C_n(T)]/T=0$, the magnitude of the specific heat jump becomes smaller compared with $\Delta_{1a}$. Moreover, the line shape of $C_s/T$ for $\Delta_{1b}$ is convex upward. 
The superconducting gap of $\Delta_{2}$ and $\Delta_{3}$ are different only by the $k^3$ term. Therefore, as is seen from Fig. \ref{fig_spe} (b), the temperature dependencies of $C_s/T$ for $\Delta_{2}$ and $\Delta_{3}$ are almost the same. In this case, $C_s/T$ has a $T^2$-behavior at low temperature since the superconducting gap has the point nodes. In the case of $\Delta_3$, as is mentioned in Sec. \ref{sec_gap}, the superconducting gap is quite similar to that for the line nodal one. For this reason, although we can see the $T^2$ behavior of $C_s/T$ very near $T/T_c=0$, the line shape of $C_s/T$ is almost the same as the line nodal one such as $\Delta_{1b}$. 

We have seen a wide variation of the temperature dependence of the specific heat for the superconducting state. However, it should be worth mentioning that $\Delta_{1a}$ and $\Delta_{1b}$ can be mixed with each other since they belong to the same irreducible representation. Hence, depending on the ratio of the mixture, the line shape of the specific heat can be similar to those of other pair potentials. 
Nevertheless, the behavior of the spin susceptibility of $\Delta_{1a}$ and $\Delta_{1b}$ is different from others, as we show below.

\subsection{Spin susceptibility}
In this section, we calculate the temperature dependence of the spin susceptibility for the possible pair potentials. From the temperature dependence of the spin susceptibility, we can see the spin structure, i.e. $d$ vector, on the Fermi surface \cite{PhysRev.131.1553,doi:10.1143/JPSJ.81.011009}. For spin-singlet superconductors, the spin susceptibility decreases with decreasing temperature for any direction and becomes zero at $T=0$. On the other hand, for spin-triplet superconductors, the temperature dependence of the spin susceptibility depends on the relation between the direction of applied magnetic field and the direction of the $d$ vector. If the magnetic field is parallel to the $d$ vector, the spin susceptibility decreases with decreasing $T$. In contrast, the magnetic field is perpendicular to the $d$ vector, the spin susceptibility does not depend on $T$. The effect of the Van Vleck susceptibility, which originates from interband scattering, is also important to understand the temperature dependence of the spin susceptibility. If the Van-Vleck effect is strong, the spin susceptibility has a finite value at $T=0$ even for the spin-singlet pairing. 

For the DSs, the Zeeman terms $h_i$ $(i=x,y,z)$, which express the coupling between electronic spin and magnetic field ${\bm H}=(H_x, H_y, H_z)$, are given by
\begin{align}
h_{x}&=\frac{1}{2}\mu_{B}H_{x}(\sigma_0+\sigma_z)s_{x},\\
h_{y}&=\frac{1}{2}\mu_{B}H_{y}(\sigma_0+\sigma_z)s_{y},\\
h_{z}&=\mu_{B}H_z\sigma_0s_{z},
\end{align}
where $\mu_B$ is the Bohr magneton and the $g$ factor is taken as 2 for simplicity. The in-plane Zeeman effect for the $p$-orbital is absent since $|p_x+ip_y\rangle$ and $|p_x-ip_y\rangle$ are orthogonal to each other.
 From the linear response theory, the spin susceptibility is given by
\begin{align}
\chi_i=-\lim_{{\bm q}\rightarrow 0}\frac{1}{N}\sum_{{\bm k},\alpha,\beta}\frac{f(E_\alpha({\bm k}))-f(E_\beta({\bm k}+{\bm q}))}{E_\alpha({\bm k})-E_\beta({\bm k}+{\bm q})}
|\langle\alpha|h_i|\beta\rangle|^2,
\end{align}
where $\alpha$ and $\beta$ are the band indices.
\begin{figure}[t]
\begin{center}
\includegraphics[width=7cm]{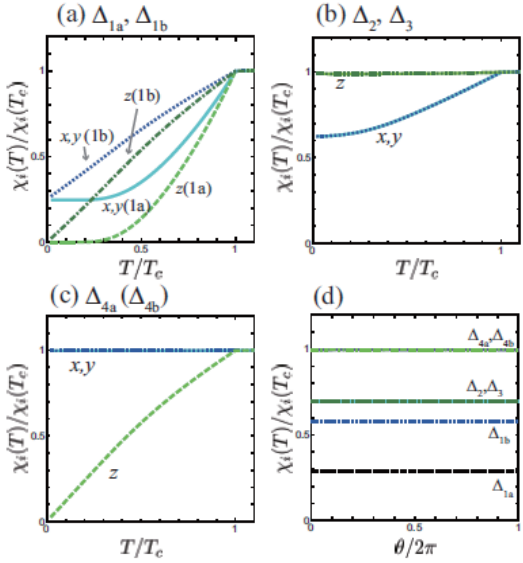}
\caption{Calculated results of the spin susceptibility for possible pair potentials. (a) Temperature dependence of the spin susceptibility for $\Delta_{1a}$ and $\Delta_{1b}$. Cyan solid line (light green dashed line) indicates $\chi_x$ and $\chi_y$ ($\chi_z$) for $\Delta_{1a}$. Blue dotted line (green dashed-dotted line) indicates $\chi_x$ and $\chi_y$ ($\chi_z$) for $\Delta_{1b}$. (b) Temperature dependence of the spin susceptibility for $\Delta_{2}$ and $\Delta_{3}$. Cyan solid line (light green solid line) indicates $\chi_x$ and $\chi_y$ ($\chi_z$) for $\Delta_{2}$. Blue dotted line (green dotted line) indicates $\chi_x$ and $\chi_y$ ($\chi_z$) for $\Delta_{3}$. (c) Temperature dependence of the spin susceptibility for $\Delta_{4a}$. Cyan solid line, blue dotted line, and light green dashed line indicate $\chi_x$, $\chi_y$ and $\chi_z$ for $\Delta_{2}$, respectively. (d) Azimuthal-angle dependence of the spin susceptibility at $T/T_c=0.4$, where $\theta/2\pi=0$ ($1/4$) indicates the $x$ ($y$) direction. Black dashed line, blue dotted line, cyan solid line, green dashed-dashed-dotted line, solid purple line and green dashed-dotted-dotted line indicates the spin susceptibility for $\Delta_{1a}$, $\Delta_{1b}$, $\Delta_{2}$, $\Delta_{3}$, $\Delta_{4a}$ and $\Delta_{4b}$, respectively.}
\label{fig_sus}
\end{center}
\end{figure}

In Fig. \ref{fig_sus} (a), we show the temperature dependence of spin susceptibility for $\Delta_{1a}$ and $\Delta_{1b}$. In the case of $\Delta_{1a}$ and $\Delta_{1b}$, the spin susceptibility for any direction decreases with decreasing $T$. However, $\chi_x$ and $\chi_y$ have finite value at $T=0$ in contrast to $\chi_z$. To examine this anisotropy, we obtain the Zeeman term in the band basis:
\begin{align}
\tilde{h}_x&
=-\frac{\eta^2}{r^2}\tilde\sigma_0(\frac{k_x^2-k_y^2}{2}s_x-k_xk_ys_y)\nonumber\\
&+\frac{m\eta^2}{r^2R}\tilde\sigma_z(\frac{k_x^2-k_y^2}{2}s_x-k_xk_ys_y)\nonumber\\
&+\frac{\eta^2}{rR}\tilde\sigma_x(\frac{k_x^2-k_y^2}{2}s_x-k_xk_ys_y),
\\
\tilde{h}_y&
=-\frac{\eta^2}{r^2}\tilde\sigma_0(k_xk_ys_x+\frac{k_x^2-k_y^2}{2}s_y)\nonumber\\
&+\frac{m\eta^2}{r^2R}\tilde\sigma_z(k_xk_ys_x+\frac{k_x^2-k_y^2}{2}s_y)\nonumber\\
&+\frac{\eta^2}{rR}\tilde\sigma_x(k_xk_ys_x+\frac{k_x^2-k_y^2}{2}s_y),
\\
\tilde{h}_z&
=\tilde\sigma_0s_z,
\end{align}
where we ignore the $k^3$ terms for simplicity. It is found that $\tilde h_x$ and $\tilde h_y$ have an interband component $\tilde \sigma_x$, which leads to the Van Vleck susceptibility. However, $\tilde h_z$ has only an intraband component. In addition, the intra-band component of $\tilde h_{x(y)}$ is $k$-dependent, while that of $\tilde h_{z}$ is not. Therefore, there is an anisotropy between $\chi_{x(y)}$ and $\chi_z$. Moreover, the inter-band component is proportional to $\eta^2$, which is a parameter related to the spin-orbit interaction. Therefore, the anisotropy can be tuned by $\eta$. It is noted that the anisotropy stemming from the spin-orbit interaction can be seen in superconducting doped TIs \cite{doi:10.7566/JPSJ.82.044704}, non-centrosymmetric superconductors \cite{doi:10.1143/JPSJ.76.034712} and locally non-centrosymmetric superconductors \cite{doi:10.1143/JPSJ.81.034702}. 
$\chi_i$ for $\Delta_{1a}$ has an exponential behavior since the superconducting gap is fully gapped. On the other hand, the line shape of $\chi_i$ for $\Delta_{1b}$ is almost linear since the superconducting gap has line nodes. Despite there are the anisotropic Van Vleck effect and the effect from the superconducting gap structure, the spin susceptibility for both $\Delta_{1a}$ and $\Delta_{1b}$ decreases enough around $T=0$.

In Fig. \ref{fig_sus} (b), we show the temperature dependence of the spin susceptibility $\chi_i$ for $\Delta_{2}$ and $\Delta_{3}$. 
As is mentioned in Sec. \ref{sec_gap}, the $d$ vector of $\Delta_{2}$ and $\Delta_{3}$ is almost parallel to $x$-$y$ plane. Therefore, $\chi_z$ is $T$ independent and $\chi_{x(y)}$ decreases with decreasing $T$. If the Van Vleck effect is absent, $\chi_{x(y)}(0)/\chi_{x(y)}(T_c)=0.5$ at $T=0$ since the $d$ vector is completely polarized in the $x$-$y$ plane.

In Fig. \ref{fig_sus} (c), we show the temperature dependence of the spin susceptibility $\chi_i$ in the case of $\Delta_{4}$. The $d$ vector for $\Delta_{4}$ is almost parallel to the $z$ axis. Consequently, $\chi_{x(y)}$ is $T$-independent and $\chi_z$ decreases with decreasing $T$. The line shape of $\chi_z$ is almost linear because the superconducting gap is a line-node-like structure. 

The azimuthal-angle dependence of the spin susceptibility at $T/T_c=0.4$ is shown in Fig. \ref{fig_sus} (d). For $\Delta_{1a}$, $\Delta_{1b}$, $\Delta_{2}$ and $\Delta_{3}$, there is no angle dependence of spin susceptibility. On the other hand, for $\Delta_{4a}$ and $\Delta_{4b}$, the $k^3$ terms lead to the in-plane anisotropy as can be seen from the analytical formula of the $d$-vector. However, due to the fact that the effect of the $k^3$ terms is small compared with the $k$-linear terms, no angle dependence is seen in the calculated results in Fig. \ref{fig_sus} (d). We note that this behavior is different from the case of the superconducting doped TIs. For superconducting doped TIs, such as Cu$_x$Bi$_2$Se$_3$, we can see an azimuthal-angle dependence in the case of $E_u$ pair potential since the $d$ vector is parallel to the $y$-$z$ or $z$-$x$ plane \cite{doi:10.7566/JPSJ.82.044704}, which has been also observed experimentally \cite{2015arXiv151207086M}. 

In summary of this subsection, we have revealed that there are wide variations of the temperature dependence of the spin susceptibility. In particular, at low temperature, the combinations of the spin susceptibility for three directions are completely different among $\Delta_1$, $\Delta_2$ ($\Delta_3$) and $\Delta_{4a}$ ($\Delta_{4b}$). It is worth mentioning that even if $\Delta_{1a}$ and $\Delta_{1b}$ are mixed, $\chi_i$ for any direction is small enough at low temperature, which is different from the cases with other pair potentials. 
In addition, we have clarified that the line shape of the temperature dependence of spin susceptibility is useful to reveal the superconductivity in the doped DSs. 
We conclude that it is possible to distinguish the superconducting states, except the difference between $\Delta_{2}$ and $\Delta_{3}$, by measuring the $x$-$y$ plane and $z$-direction Knight shift.

\section{Surface physical properties}\label{sec_surface}
It has been well known that unconventional superconductors host exotic SABSs \cite{PhysRevLett.74.3451,0034-4885-63-10-202,PhysRevB.56.7847,PhysRevLett.107.077003}. 
In this section, we reveal the SABSs of the superconducting DSs using the recursive Green's function method \cite{PhysRevB.55.5266}. We also interpret the calculated results with topological numbers. 

First of all, we briefly review the surface state in the normal state \cite{PhysRevB.85.195320}. In Fig. \ref{fig_surface_n}, we show the surface Brillouin zone and spectral function for the (010) and (110) surface. For the DSs, the mirror Chern number with respect to the $x$-$y$ mirror plane is 1. Here, the mirror Chern number is defined as
\begin{align}
n_{\cal M}=\frac{n_{+i}-n_{-i}}{2},
\end{align}
where $n_{\pm i}$ is the Chern number for mirror subsectors labeled with mirror eigenvalues $\pm i$. 
The non-zero mirror Chern number ensures the existence of surface states as shown in Figs. \ref{fig_surface_n} (b) and \ref{fig_surface_n} (c). 
Along the directions of $\bar \Gamma_1$-$\bar Z_1$ and $\bar \Gamma_2$-$\bar Z_2$, the surface state has the form connecting the $\bar{\Gamma}$ and the bulk Dirac point. 
On the other hand, along the $\bar \Gamma_1$-$\bar X_1$ and $\bar \Gamma_2$-$\bar X_2$ directions, the surface state looks like a cone shape. The spectral functions at $E=0.07$ eV are shown in Figs. \ref{fig_surface_n} (d) and (e). At $E=0.07$ eV, there are both bulk and surface states.

\subsection{Surface Andreev bound state}
\begin{figure}[t]
\begin{center}
\includegraphics[width=8cm]{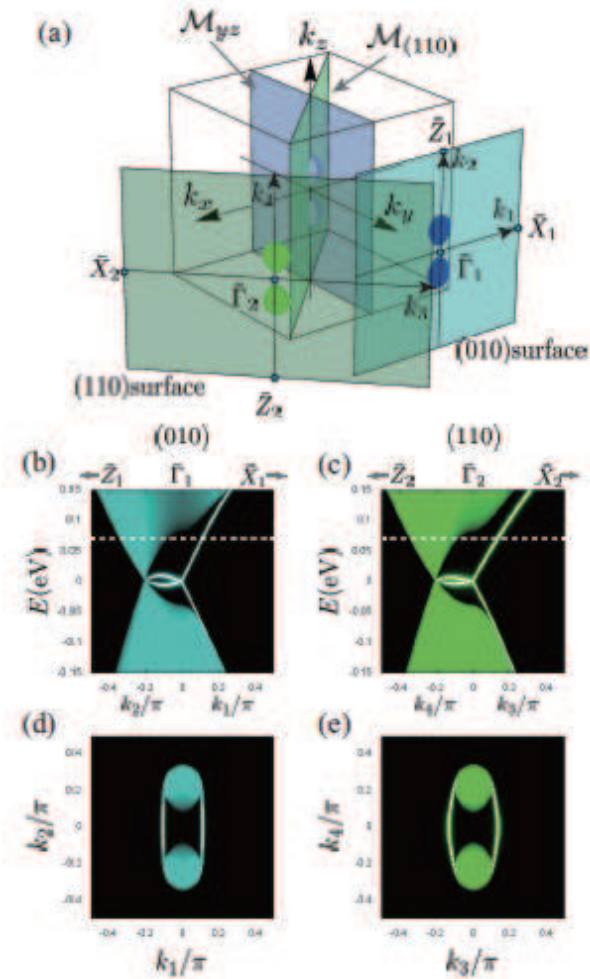}
\caption{(a) (010) and (110) surface Brillouin zones, and the projected Fermi surface. (b) [(c)] Surface spectral function of the normal state for the (010) [(110)] surface along the $\bar Z_1$-$\bar \Gamma_1$-$\bar X_1$ ($\bar Z_2$-$\bar \Gamma_2$-$\bar X_2$) line. The isolated branches indicate the surface states. White dotted lines indicate $E=0.07$ eV. (d) [(e)] Surface spectral function of the normal state for the (010) [(110)] surface at $E=0.07$ eV.}
\label{fig_surface_n}
\end{center}
\end{figure}
\begin{figure*}[t]
\begin{center}
\includegraphics[width=16cm]{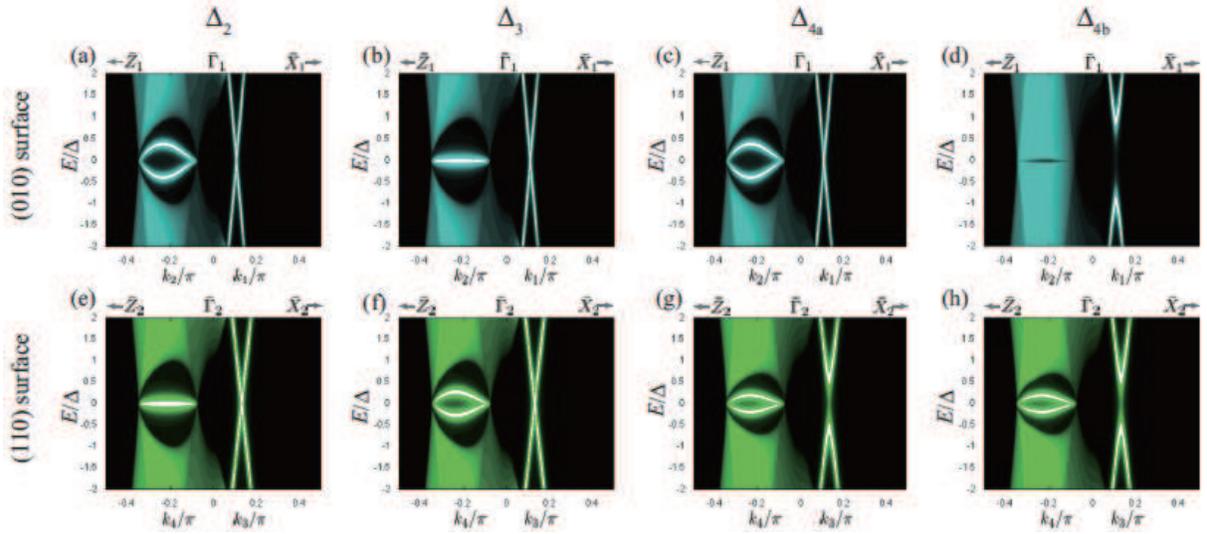}
\caption{Surface Andreev bound state on the (010) (a)-(d) and (110) (e)-(h) surface for the possible pair potentials.}
\label{fig_surface}
\end{center}
\end{figure*}
In Fig. \ref{fig_surface}, we show our calculated results of surface spectral function for the (010) and (110) surfaces in the superconducting state. Here, we focus on $\Delta_2$, $\Delta_3$ and $\Delta_{4a}$ ($\Delta_{4b}$) pairings and ignore $\Delta_1$ pairings since there is no SABS in the case of $\Delta_1$. It is noted that, in the case of $\Delta_4$, the (010) surface for $\Delta_{4a}$ corresponds to the (100) surface for $\Delta_{4b}$, and vice versa, since $\Delta_{4}$ is the two-dimensional representation. 

First, we would like to mention the surface state along the $\bar\Gamma_1$-$\bar Z_1$ ($\bar\Gamma_2$-$\bar Z_2$) direction for the (010) [(110)] surface.
As can be seen from the upper column of Fig. \ref{fig_surface}, $\Delta_{2}$ and $\Delta_{4a}$ have dispersive SABSs and $\Delta_{3}$ has flat SABSs along the $\bar{\Gamma}_1$-$\bar{Z}_1$ direction between the point nodes. 
On the other hand, as is seen from the lower column of Fig. \ref{fig_surface}, $\Delta_{2}$ has flat SABSs, and $\Delta_{3}$, $\Delta_{4a}$, and $\Delta_{4b}$ have dispersive SABSs along the $\bar{\Gamma}_2$-$\bar{Z}_2$ line between the point nodes. 

To interpret the surface states, we see the topological nature of the system. 
Since the unconventional pair potentials, $\Delta_2$, $\Delta_3$ and $\Delta_4$, support nodes in the bulk superconducting gaps, bulk topological numbers are defined within lower dimensional subspaces in the Brillouin zone. In time-reversal-invariant odd-parity superconductors, the stability of point nodes is ensured by a one-dimensional winding number with the aid of mirror-reflection symmetry \cite{PhysRevB.90.024516}. From the topological classification of nodes, the nontrivial winding number exists only if the pair potential is even under the mirror-reflection operation and predicts flat SABSs between the point nodes as long as we make the surface perpendicular to the mirror invariant plane. Note that, although the point nodes in the DS with $\Delta_{2}$, $\Delta_{3}$, $\Delta_{4a}$, and $\Delta_{4b}$ are protected by the $C_4$ rotational symmetry, it is necessary to consider mirror-reflection symmetry to understand flat SABSs. This is because the $C_4$ rotational symmetry cannot survive when we make the surface normal to the rotational axis. 

In the case of the DS, there are two mirror planes parallel to $z$ axis, i.e., $y$-$z$ ($z$-$x$) plane and the (110) plane. 
As is summarized in Table \ref{table_pair}, 
$\Delta_{2}$ is even under the (110) mirror operation but odd under the $y$-$z$ mirror operation. Conversely, $\Delta_{3}$ is even under the $y$-$z$ mirror operation but odd under the (110) mirror operation. 
On the other hand, $\Delta_{4a}$ ($\Delta_{4b}$) is even under the $y$-$z$ ($z$-$x$) mirror operation but odd under the $z$-$x$ ($y$-$z$) mirror operation and does not have the (110) mirror-reflection symmetry. 

\begin{figure}[t]
\begin{center}
\includegraphics[width=8cm]{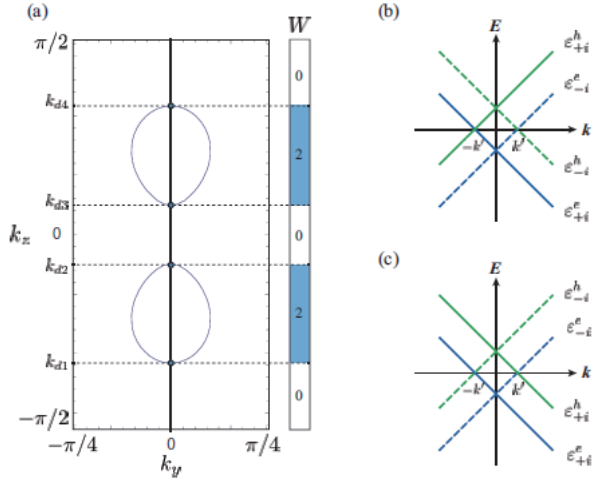}
\caption{(a) Fermi surface on the $y$-$z$ mirror plane and one-dimensional winding number $W$ for $\Delta_{3}$ at a fixed $k_z$. Blue dots indicate the position of the point node at $k_z=k_{di}$ $(i=1,2,3,4)$. For $k_{d1}<k_z<k_{d2}$  and $k_{d3}<k_z<k_{d4}$, the winding number takes $W=2$, and otherwise $W=0$. (b) [(c)] Schematic picture of surface electron and hole band when the parity of pair potential under the mirror operation is even (odd).}
\label{fig_winding}
\end{center}
\end{figure}

To examine the stability of the flat SABSs, we numerically calculate the one-dimensional winding number on the mirror-even plane for each pair potential. Here, we focus on the $y$-$z$ plane in the case of $\Delta_{3}$. For the BdG Hamiltonian, the $y$-$z$ mirror operator, time-reversal operator, and particle-hole operator are given by 
\begin{align}
{\cal M}_{yz}^{\prime}&=
\begin{pmatrix}
{\cal M}_{yz}&0\\
0&{\cal M}_{yz}^*
\end{pmatrix},\\
{\cal T}^{\prime}&=
{\cal T}\tau_0
,\\
{\cal C}&=
\sigma_0 s_0\tau_x\cal K.
\end{align}
Combining these operators, we can define the mirror-chiral operator:
\begin{align}
\Gamma_{\cal M}={\cal M}_{yz}^{\prime} {\cal T}^{\prime} {\cal C},
\end{align}
which satisfies
\begin{align}
\Gamma_{\cal M}H_{\rm BdG}(k_x, k_y, k_z)\Gamma_{\cal M}^\dagger
=-H_{\rm BdG}(-k_x, k_y, k_z).
\end{align}
With this mirror-chiral operator, we can define the one-dimensional winding number for a fixed $k_z$:
\begin{align}
W=-\frac{1}{4\pi i}\int_{-\pi}^\pi dk_y {\rm Tr} (\Gamma_{\cal M} H_{\rm BdG}^{-1}\partial_{k_y}H_{\rm BdG})\label{winding}.
\end{align}
To evaluate Eq. (\ref{winding}), we follow the method in Ref. \cite{PhysRevB.83.224511}. 
 As a result, we find that the winding number $W$ is 
\begin{align}
W=\begin{cases}
   2\hspace{10pt}(k_{d1}<k_z<k_{d2},k_{d3}<k_z<k_{d4})\\
   0\hspace{10pt}{\rm otherwise} 
  \end{cases},
\end{align}
where $k_{di} (i=1,2,3,4)$ are the momenta of the disconnected Fermi surfaces around the Dirac points as shown in Fig. \ref{fig_winding}. Therefore, the existence of the zero-energy flat SABSs at $k_{d1}<k_z<k_{d2}$ and $k_{d3}<k_z<k_{d4}$ is ensured by the non-zero winding number. In the same manner, we can easily interpret the zero-energy flat SABSs for the (110) surface in the case of $\Delta_{2}$ with the winding number. 

In the case of $\Delta_{4a}$, the gap function is odd under the $y$-$z$ mirror operation, and hence there are dispersive SABSs between the point nodes along the $\bar\Gamma_1$-$\bar Z_1$ direction for the (010) surface. On the other hand, in the case of $\Delta_{4b}$, the gap function is even under the $y$-$z$ mirror operation, where we can define the one-dimensional winding number. However, the winding number is zero for any $k_z$. Correspondingly, there is no SABS along the $\bar\Gamma_1$-$\bar Z_1$ direction as shown in Fig. \ref{fig_surface} (d). For the (110) surface, both $\Delta_{4a}$ and $\Delta_{4b}$ do not have mirror-reflection symmetry, hosting the dispersive SABSs between the point nodes. 

Next, we refer to the surface state along the $\bar\Gamma_1$-$\bar X_1$ ($\bar\Gamma_2$-$\bar X_2$) direction for the (010) [(110)] surface. As is seen from Figs. \ref{fig_surface_n} (d) and \ref{fig_surface_n} (e), there are surface states but no bulk states at the Fermi level in the normal state. In the superconducting state, as shown in Fig. \ref{fig_surface}, there are zero-energy SABSs for the (010) surface in the case of $\Delta_{2}$, $\Delta_{3}$, and $\Delta_{4a}$, and for the (110) surface in the case of $\Delta_{2}$ and $\Delta_{3}$. These zero-energy SABSs are formed by the remaining surface states in the normal state. Hereafter, we call this type of zero-energy SABS as remaining zero-energy SABS. 

 We can understand whether the normal surface states remain gapless or not in the superconducting state with the parity of the pair potential under the mirror operation ${\cal M}_{xy}$ \cite{PhysRevLett.111.087002}. In the superconducting state, the hole band $\varepsilon^h(k)$ appears, where the relation between electron band and hole band is $\varepsilon^h(k)=-\varepsilon^e(-k)$. If the pair potential is even (odd) under the mirror operation ${\cal M}_{xy}$, the BdG Hamiltonian commutes with a mirror operator ${\cal M}_{xy}^{\eta_{\cal M}}={\rm diag}[{\cal M}_{xy},\eta_{\cal M}{\cal M}^*_{xy}]$ where $\eta_{\cal M}=+$ $(-)$. Then the BdG Hamiltonian can be block diagonalized by the eigenvectors of ${\cal M}_{xy}^{\eta_{\cal M}}$, where each block diagonalized part is labeled with mirror eigenvalue, $+i$ or $-i$. If the parity of the pair potential under the mirror operation is even, the particle-hole symmetry exists between the different mirror eigensectors, which can be schematically shown in Fig. \ref{fig_winding} (b). In this case, the band with the same mirror eigensector crosses at $E=0$, namely, $\varepsilon_{+i}^e(-k^\prime)=\varepsilon_{+i}^h(-k^\prime)$ and $\varepsilon_{-i}^e(k^\prime)=\varepsilon_{-i}^h(k^\prime)$ at $E=0$, and hence it becomes gapped. On the other hand, if the parity of the pair potential under the mirror operation is odd, each mirror eigensector has the particle-hole symmetry. Then, the bands with opposite mirror eigenvalues cross each other at $E=0$, namely, $\varepsilon_{+i}^e(-k^\prime)=\varepsilon_{-i}^h(-k^\prime)$ and $\varepsilon_{-i}^e(k^\prime)=\varepsilon_{+i}^h(k^\prime)$ at $E=0$ as shown in Fig. \ref{fig_winding} (b). In this case, the crossed bands cannot mix with each other and remain gapless. For these reasons, the remaining zero-energy SABSs appear on the $\bar\Gamma_1$-$\bar X_1$ and $\bar\Gamma_2$-$\bar X_2$ line in the case of $\Delta_{2}$ and $\Delta_{3}$. It should be noted that the mirror Chern number $n^\prime_{\cal M}$ defined with ${\cal M}_{xy}^-$  is 2 in the case of $\Delta_{2}$ and $\Delta_{3}$, which corresponds to the zero-energy states. 

The remaining zero-energy SABS on the $\bar \Gamma_1$-$\bar X_1$ for the (010) surface in the case of $\Delta_{4a}$ is protected by a different reason. In this case, the zero-energy state can be interpreted with a zero-dimensional topological number that is defined by combining the $y$-$z$ mirror operator ${\cal M}_{yz}^\prime$ and the chiral operator ${\cal T}^{\prime} {\cal C}$ \cite{PhysRevLett.115.187001}. The zero-dimensional topological number for each $k_x$ is given by
 
\begin{align}
\rho(k_x)={\rm sgn} \left\{{\rm Pf}[H^{\rm SS}_{\rm BdG}(0,k_x)is_z\tau_x] \right\},\label{0DTN}
\end{align}
with the surface BdG Hamiltoninan and surface normal Hamiltonian:
\begin{align}
H^{\rm SS}_{\rm BdG}(k_z,k_x)&=[H^{\rm SS}(k_z,k_x)-\mu]\tau_z+\Delta^{\rm surf}\tau_x,\\
H^{\rm SS}(k_z,k_x)&=v(k_zs_x-k_xs_z).
\end{align}
By changing $k_x$, $\rho(k_x)$ changes its sign when $k_x$ passes through the momentum of the zero-energy surface state, which means that the surface state is protected topologically. This zero-dimensional topological number can be defined when the parity of the pair potentials under the relevant mirror operation is odd. Therefore, we can also explain the remaining zero-energy SABS on the $\bar \Gamma_1$-$\bar X_1$ for the (010) surface in the case of $\Delta_{2}$ [the $\bar \Gamma_2$-$\bar X_2$ for the (110) surface in the case of $\Delta_{3}$] with the zero-dimensional topological number for the $y$-$z$ [(110)] mirror plane in addition to the mirror Chern number for the $x$-$y$ mirror plane.
For more details and derivation of the zero-dimensional topological number $\rho(k_x)$, see the supplemental material in Ref. \cite{PhysRevLett.115.187001}. 

Here, we make a comment on the previous work done by two of the present authors. In the previous work, the minimal required symmetries of the DSs, namely, time-reversal, inversion, and four fold rotational symmetries, have been assumed. Under this assumption, surface Majorana quartet has been predicted 
when four fold rotation symmetry is broken into two fold one.
On the other hand, 
in the present paper, full $D_{4h}$ crystal symmetries have been taken into account,  which enables us to realize different surface states such as the flat surface state. It should be noted that the Majorana quartet appears even in our model in the case of $\Delta_{2}$ and $\Delta_{3}$ by breaking the four fold rotational symmetry.

In summary of this section, we have found that wide variation of exotic SABSs appear depending on the pair potentials and surface direction. We have revealed that the zero-energy SABSs can be interpreted with three topological numbers: one-dimensional winding number, two-dimensional-mirror Chern number, and zero-dimensional topological number. The mirror-reflection symmetry plays an essential role to protect SABSs in this system. The zero-energy SABS and relevant topological number are summarized in Table \ref{TN}. 
\begin{table}[b]
\begin{center}
\begin{tabular}{ccc}
\hline\hline
Topo. No.&zero-energy SABS&pair potential and surface\\
\hline
$W$&flat&$\Delta_2$(110),$\Delta_{3}$(010)\\
$n_{\cal M}^\prime$&remaining&$\Delta_2$[(010),(110)],$\Delta_3$[(010),(110)]\\
$\rho(k_x)$&remaining&$\Delta_2$(010),$\Delta_3$(110),$\Delta_{4a}$(010)\\
\hline\hline
\end{tabular}
\caption{Zero-energy SABS and relevant topological number. First column shows the type of topological number: one-dimensional winding number $W$, mirror Chern number $n_{\cal M}^\prime$ and zero-dimensional topological number $\rho(k_x)$. Second column shows the type of zero-energy SABS and the third one shows the pair potentials and surface orientations that host the zero-energy SABS characterized with the topological number.}
\label{TN}
\end{center}
\end{table}
\subsection{Surface density of state}
\begin{figure}[t]
\begin{center}
\includegraphics[width=5cm]{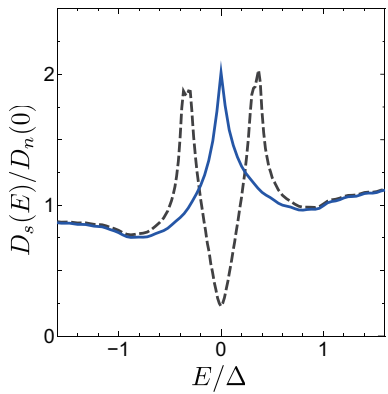}
\caption{SDOS as a function of normalized energy for the (010) surface. Black dashed line (blue solid line) indicates the SDOS for $\Delta_{2}$ ($\Delta_{3}$) which corresponds to the dispersive (flat) SABSs case.}
\label{fig_SDOS}
\end{center}
\end{figure}
In this section, we calculate the surface density of state (SDOS) by integrating the spectral function with respect to the wave number $k$ obtained in the previous section. In Fig. \ref{fig_SDOS}, we show the normalized SDOS $D_s(E)/D_n(0)$ as a function of normalized energy $E/\Delta$ for the (010) surface in the case of $\Delta_{2}$ and $\Delta_{3}$, which corresponds to the case of the dispersive SABSs and flat SABSs, respectively. In the case of dispersive SABSs, double peaks appear around $E/\Delta$=0.5 and the spectra look like V shape. The double peaks originate from saddle points on the top (bottom) of the arc in $E>0$ ($E<0$). On the other hand, the flat SABS leads to a single peak at $E/\Delta=0$ in the SDOS. 
These results suggest that we can distinguish $\Delta_{2}$ and $\Delta_{3}$ by conductance measurement with low transmissivity realized in scanning tunneling spectroscopy. 

We briefly mention the experimental results of point contact spectra for Cd$_3$As$_2$ \cite{10.1038/nmat4455,10.1038/nmat4456}. The experiments have been done for the (112) natural cleavage surface and a zero-energy conductance peak has been observed. In general, one of the promising explanations of a zero-energy conductance peak in the superconducting state is the existence of a zero-energy flat band. 
When we see the (112) surface of Cd$_3$As$_2$, the (110) mirror-reflection symmetry is preserved. Therefore, $\Delta_{2}$ that has a flat band for this direction can explain the observed zero-energy conductance peak. 
Besides the topologically protected flat band, the accidental flat band stemming from the remaining normal surface state also leads to a zero-energy conductance peak. 
It has been revealed that, in the superconducting doped TIs with fully gapped odd-parity pairing, the dispersion of SABSs has a structural transition from a simple cone shape to a twisted shape \cite{PhysRevB.85.180509,PhysRevB.83.134516,PhysRevLett.108.107005}. At the transition point, a zero-energy flat-like band appears, leading to the zero-energy conductance peak. In the case of the superconducting DSs, the SABSs exist between the point nodes along $\Gamma_2$-$Z_2$. Therefore, as long as the point nodes are protected, the SABSs cannot mix with the remaining surface state, which means the SABSs cannot change their structure. 
For further discussion of the experiments, it is necessary to calculate the conductance with high transmissivity. 

\section{Discussion}\label{sec_disc}
So far, we have revealed the superconductivity in DSs. Here, we briefly mention the difference between superconductivity in the DSs and that in Bi$_2$Se$_3$ type TIs focusing on the superconducting gap structure and the symmetry of the pair potential. In Table \ref{table_TI_DS}, we show the type of superconducting gap structure on the Fermi surface and irreducible representation of the DSs and TIs, assuming the $D_{4h}$ crystal. The details of the superconducting gap structure of TIs are shown in Appendix. As is obvious from Table \ref{table_TI_DS}, the superconducting gap structure is totally different between DSs and TIs though the matrix forms of the pair potentials are the same. In particular, different from the superconductivity in the TIs, that in the DSs cannot lead to a fully-gapped topological superconductor. This difference originates from the difference of effective orbitals and topological nature.
\begin{table}[t]
\begin{center}
\begin{tabular}{ccc}
\hline\hline
$\Delta$& Dirac semimetal&topological insulator\\
\hline
$\Delta\sigma_0s_0$&FG($A_{1g}$)&FG($A_{1g}$)\\
$\Delta\sigma_zs_0$&PN($A_{1g}$)&LN($A_{2u}$)\\
$\Delta\sigma_ys_y$&PN($B_{1u}$)&PN($E_{u}$)\\
$\Delta\sigma_ys_x$&PN($B_{2u}$)&PN($E_{u}$)\\
$\Delta\sigma_xs_0$&FG($E_{u}$)&LN($A_{1g}$)\\
$\Delta\sigma_ys_z$&FG($E_{u}$)&LN($A_{1u}$)\\
\hline\hline
\end{tabular}
\caption{Possible pair potentials for two orbital models. Superconducting gap structures projected onto the conduction band (the effect of the $k^3$ terms are ignored) of the Dirac semimetal and topological insulator and the irreducible representation classified as the $D_{4h}$ point group are shown. FG, PN, and LN stand for full gap, point node, and line node, respectively.}
\label{table_TI_DS}
\end{center}
\end{table}

In this paper, we have focused on Cd$_3$As$_2$ class DSs where the bulk Dirac points are protected by a four fold rotational symmetry and effective orbitals are formed by $s$- and $p$-orbitals. Pair potentials in the band basis for DSs protected by other discrete rotational symmetry can be easily obtained by changing the basis functions Eqs. (\ref{eq_a})-(\ref{eq_e}) in Eqs. (\ref{eq_d1a})-(\ref{eq_d4b}). Then, we can interpret the superconducting gap structure and $d$ vector. 
However, it should be noted that we have to reconsider the surface states and the stability of nodes carefully for other DSs since the topological nature can be changed depending on the related discrete rotational symmetry. 
If the Dirac points are made of $s$ and $d$ orbitals, different types of DSs are realized \cite{10.1038/ncomms5898}. It would be interesting to explore the superconductivity in this type of DSs since the $d$-wave superconductivity is realized \cite{sddirac}.

\section{Summary}\label{sec_summary}
We have studied the superconductivity in Cd$_3$As$_2$ type DSs. By obtaining a single-band description of the pair potentials, we have clarified the superconducting gap structure and $d$ vector on the Fermi surface. It has been found that the superconducting gap structure can be classified into four types: isotropic full gap, point node at poles, horizontal line node and vertical line node (effectively) . For the spin-triplet case, the direction of the $d$ vector is classified into two types: $d$ $\perp$ $z$-axis and $d$ $\parallel$ $z$-axis. These characteristics of the superconducting states are completely different from those of the superconducting TIs and other topological materials. By solving the linearized gap equation, we have found that the nodal spin-triplet pairing ($\Delta_{2}$ and $\Delta_{3}$) can be stabilized when the interorbital attraction is sufficiently stronger than the intraorbital one. Moreover, we have revealed that orbit-momentum locking in the DS plays a key role to interpret the superconducting gap structure of the possible superconducting state. We have also calculated the physical properties of each pair potential. Our calculation results of the bulk physical properties, especially the spin susceptibility, are useful to reveal the pairing symmetry. Moreover, we have shown that exotic SABSs exist in $\Delta_2$, $\Delta_3$, $\Delta_{4a}$, and $\Delta_{4b}$, and interpreted each SABS with three types of topological numbers. We have concluded that it is possible to distinguish the possible superconducting states by combining the bulk and surface measurements. 
 
\section{Acknowledgements}
We thank Takeshi Mizushima, Xiong-Jun Liu, Keiji Yada and Ai Yamakage for valuable discussions. 
This work was supported by the ``Topological Materials Science" Grant-in Aid for Scientific Research on Innovative Areas from the Ministry of Education, Culture, Sports, Science and Technology of Japan (No. 15H05853, 15H05855), Grant-in-aid for JSPS Fellows (No. 26010542) (T. H.), Grant-in-Aid for Scientific Research B (No. 15H03686) (Y. T.) and (No. 25287085)(M. S.), and  Grant-in-Aid for Challenging Exploratory Research (No. 15K13498) (Y. T.). 

\appendix

\section{Difference between superconductivity in Dirac semimetals and topological insulators}
The low energy electronic state of three-dimensional topological insulator Bi$_2$Se$_3$ is described by 4 $\times$ 4 Dirac Hamiltonian \cite{PhysRevB.82.045122}:
\begin{align}
H_{\rm TI}({\bm k})=m\sigma_x+v_zk_z\sigma_y+v(k_xs_y-k_ys_x)\sigma_z,
\end{align}
where $\sigma_i$ and $s_i$ $(i=0,x,y,z)$ are the Pauli matrices in the orbital and spin space and the effective orbitals consist of two $p_z$ orbitals.
For this model, we can consider the six types of $k$-independent pair potentials: $\Delta\sigma_0s_0$, $\Delta\sigma_zs_0$, $\Delta\sigma_ys_y$, $\Delta\sigma_ys_x$, $\Delta\sigma_xs_0$ and $\Delta\sigma_ys_z$  \cite{PhysRevLett.105.097001}. As we have obtained in Ref. \cite{doi:10.7566/JPSJ.82.044704}, the possible pair potentials projected onto the conduction band are 
\begin{align}
\Delta\sigma_0s_0&\rightarrow \Delta s_0,\label{eq_TI1}\\
\Delta\sigma_zs_0&\rightarrow \Delta \frac{m}{\lambda} s_0,\\
\Delta\sigma_ys_y&\rightarrow\Delta\left[ 
v_zk_z\left(\frac{k_xk_y}{\nu k_\parallel}-\frac{k_xk_y}{\lambda k_\parallel}\right)s_x
\right . \nonumber\\
&\left . +v_zk_z\left(\frac{k_y^2}{\nu k_\parallel^2}+\frac{k_x^2}{\lambda k_\parallel^2}\right)s_y
- \frac{m v k_y}{\nu \lambda}s_z \right],\\
\Delta\sigma_ys_x&\rightarrow\Delta\left[ v_zk_z\left(\frac{k_x^2}{\nu k_\parallel^2}+\frac{k_y^2}{\lambda k_\parallel^2}\right)s_x
\right . \nonumber\\
&\left .
+ v_zk_z\left(\frac{k_xk_y}{\nu k_\parallel}-\frac{k_xk_y}{\lambda k_\parallel}\right)s_y 
- \frac{m v k_x}{\nu \lambda}s_z \right],\\
\Delta\sigma_xs_0&\rightarrow \Delta \frac{v}{\lambda}\left(k_xs_y - k_ys_x\right),\\
\Delta\sigma_ys_z&\rightarrow \Delta \left( \frac{m v k_x}{\nu\lambda}s_x +\frac{m v k_y}{\nu\lambda}s_y + \frac{v_z k_z}{\lambda}s_z\right),\label{eq_TI2}
\end{align}
where $k_\parallel=\sqrt{k_x^2+k_y^2}$, $\nu=\sqrt{m^2+v_z^2k_z^2}$, and $\lambda=\sqrt{m^2+v^2k_\parallel^2+v_z^2k_z^2}$. In the case of the TIs, two fully gapped spin-singlet pairings, one fully gapped spin-triplet pairing,
 and three point-nodal spin-triplet pairings (point nodes exist on different axis, $k_x$, $k_y$ or $k_z$) can appear, which are completely different from the superconductivity in the DSs as summarized in Table \ref{table_TI_DS} in the main text. 
\bibliography{ref}

\begin{thebibliography}{100}
\bibitem[{foo()}]{footnote1}
\bibinfo{note}{Note that if one changes the basis as
  $U_{\sigma\sigma}c_{\sigma}$, the gap function $\Delta_i$ changes as
  $U\Delta_i U^t$}.
\end{thebibliography}
\end{document}